\documentclass[12pt]{article}
\usepackage{amsbsy}

\usepackage{amssymb}

\let\vp=\varphi

\def \del {\partial}
\def \tg {\tilde \gamma}

\def \s {\sigma}
\def \ov {\over}
\def \td {\tilde}

\def \tr {{\rm \tr}}

\newcommand{\be}{\begin{equation}}
\newcommand{\ee}{\end{equation}}
\newcommand{\bea}{\begin{eqnarray}}
\newcommand{\eea}{\end{eqnarray}}

\def \la{\label}
\newcommand{\rf}[1]{(\ref{#1})}

\def\ov{\over}

\def\L{{\Lambda}}

\def\bfsigma{{\boldsymbol{\sigma}}}
\def\bfa{{\boldsymbol{\alpha}}}

\def \cU {{\cal U}}

\def \ha {{1\ov 2}}

\def \adss{$AdS_5 \times S^5$\ }

\def\J{{\cal J}}
\def\N{{\cal N}}
\def\a{\alpha}
\def\b{\beta}

\def\l{\lambda}
\def\tl{\tilde{\l}}

\def\s{\sigma}

\def\g{\gamma}
\def\tg{\tilde{\gamma}}
\def\e{\epsilon}
\def\la{\label}
\def\={\, =\, }
\def\+{\, +\, }
\def\-{\, -\, }

\def\r{\rho}
\def\k{\kappa}

\def\={\, =\, }
\def\+{\, +\, }
\def\-{\, -\, }

\def\p{\phi}
\def\ttp{\,\widetilde{\!{\widetilde{\phi}}}}

\def \pa{\partial}
\def \Tr {{\rm Tr}}

\def \H{{\cal H}}
\def\rrangle{{\rangle\!\rangle}}
\def\llangle{{\langle\!\langle}}

\def\ov{\over}
\def \ha { { 1 \over 2}}

\def \JJ {{\cal I}}

\def \del {\pa}

\def\IC{\mathbb{C}}
\def\IZ{\mathbb{Z}}
\def\IP{\mathbb{P}}

\def\id{\mathbb{I}}

\def\ialf{{\textstyle{\frac{i}{2}}}}

\def\bfgamma{{\boldsymbol{\gamma}}}

\def \sql {\sqrt{\lambda}}\def \E {{\cal E}}

\def \YM {{\rm YM}}

\def \N {{\cal N}}
\def \ci{\cite}

\def \adss {${AdS}_5 \times {S}^5$\ }
\def \ci{\cite}
\def \foot {\footnote}
\def \N {{\cal N}}
\def \l {\lambda}

\def \bi {\bibitem}

\def \th {\theta}\def \r {\rho}

\def \d {\partial}

\def \ha {{\textstyle{1 \ov 2}}}

\def \J {{\cal J}}

\def \gga {{  \bfgamma}}
\def \tgg {\td {  \gga}}

\def \D {\Delta}

\newcommand{\uu}[3]{u_{{#1}, {#3}}^{(#2)}}

\newcommand{\xx}[3]{x_{{#1}, {#3}}^{(#2)}}

\newcommand{\sw}[3]{w_{{#1}, {#3}}^{(#2)}}

\makeatletter
\renewcommand\section{\@startsection {section}{1}{\z@}%
                                   {-3.5ex \@plus -1ex \@minus -.2ex}%
                                   {2.3ex \@plus.2ex}%
                                   {\normalfont\large\bfseries}}

\renewcommand\subsection{\@startsection{subsection}{2}{\z@}%
                                   {-3.25ex\@plus -1ex \@minus -.2ex}%
                                   {1.5ex \@plus .2ex}%
       {\normalfont\normalsize\bfseries}}
\makeatother

\begin{document}

\def \tg {{\tilde \g}}
\def \ep {\epsilon}

\def \sumi {\sum^3_{i=1}}
\def \ta {\td \a}
\def \L {{\cal L}}
\def \d {\partial}

\def \H {{\cal H}}

\def \JJ {{\rm  J}}
\def \G {\Gamma}

\def \const {{\rm const}}
\def \La {\Lambda}

\def \k {\kappa}
\def \H {{\cal H}}
\def \bg {{\bar \g}}
\def \bgg {{\bar \gga}}

\def \bk {{\bar \bfsigma}}
\def \tk {{\tilde \bfsigma}}

\def \thr {{1 \ov 3}}

\def \cP {{\cal P}}

\def \sumn{\sum^3_{m,n=1}}
\def \P {\Phi}

\def \ppp { \P_1^{J_1} \P_2^{J_2}  \P_3^{J_3}   }

\def \eps {\epsilon}

\def \sumn {\sum_{m,n=1}^3}
\def \hh {{\rm h}}

\def \ta {\tilde \psi}

%%%%%%%%%%%%%%%%%%%%%%%%%%%%%%%%%
%%%
%%%%%%%%%%%%%%%%%%%%%%%%%%%%%%%%%

\begin{titlepage}

\begin{flushleft}
\hfill PUPT-2167\\
\hfill AEI-2005-118
\end{flushleft}
\vspace{-1cm}

\begin{flushleft}
\end{flushleft}

%\vspace{-2cm}

\bigskip

\vspace{1cm}
\begin{center}
{\Large\bf
Gauge-string duality  for  (non)supersymmetric  deformations of
\vspace{0.4cm}
 N=4 Super Yang-Mills theory
}

\vspace{0.3cm}

\vspace{.5cm} {S.A.~Frolov$^{a,}$\footnote{Also at SUNYIT, Utica, USA,
and Steklov 
Mathematical Institute, Moscow. frolovs@aei.mpg.de},
R.~Roiban$^{b,}$\footnote{rroiban@princeton.edu} and
A.A.~Tseytlin$^{c,}\footnote{Also at Imperial College London
and  Lebedev  Institute, Moscow. tseytlin@mps.ohio-state.edu}$}\\
\vskip 0.3cm

{\em $^{a}$Max-Planck-Institute for Gravitational Physics,
Albert-Einstein-Institute,\\
Am M\"uhlenberg 1, D-14476 Golm, Germany\\
\vskip 0.1cm
$^{b}$Department of Physics, Princeton University,\\
 Princeton, NJ 08544, USA\\
\vskip 0.1cm
$^{c}$Department of Physics, The Ohio State University,\\
 Columbus, OH 43210, USA
     }

\end{center}

\vspace{0.2cm}

\begin{abstract}
We consider a non-supersymmetric example of the AdS/CFT  duality
which generalizes the supersymmetric exactly marginal deformation
constructed in hep-th/0502086.
The string theory background we use  was
found  in  hep-th/0503201  from the $AdS_5 \times S^5$
 by a combination of T-dualities and
shifts of angular coordinates. It depends on
  three real parameters $ \g_i$
which determine the shape of the deformed 5-sphere.
The dual gauge theory has the same field content as $\N=4$ SYM theory,
but with scalar  and Yukawa interactions  ``deformed'' by $\g_i$-dependent
phases. The  special case of equal $\g_i=\g$ 
 corresponds to the  $\N=1$
supersymmetric deformation.
We  compare the energies of  semiclassical strings
 with three large  angular momenta
to the 1-loop anomalous dimensions  of the corresponding gauge-theory
scalar  operators and find that they match as it was the case
in the $SU(3)$ sector of the standard  AdS/CFT duality.
In the supersymmetric  case of equal $\g_i$
this extends the result of our previous work (hep-th/0503192)
from the 2-spin  to the 3-spin sector. This  extension turns out to be
quite nontrivial. To match the corresponding low-energy effective
``Landau-Lifshitz'' actions  on the string theory and the
gauge theory sides one  is to make a special choice of the spin chain
Hamiltonian  representing the 1-loop gauge theory dilatation operator.
This choice is adapted  to low-energy approximation, i.e. it
allows one to capture the right vacuum states and the  ``macroscopic spin wave''
sector of states of the spin chain in the continuum coherent state
effective action.
\end{abstract}

\end{titlepage}

\setcounter{footnote}{0}

\renewcommand{\theequation}{1.\arabic{equation}}
 \setcounter{equation}{0}

%%%%%%%%%%%%%%%%%%%%%%%%%%%%%%%%%
\section{Introduction}
%%%%%%%%%%%%%%%%%%%%%%%%%%%%

 Study of  AdS/CFT duality
in situations  with reduced (or no)  supersymmetry is of
obvious interest and importance. Recently,  a new
example of such duality  between an exactly marginal (in 4d sense)
deformation of $\N=4$  super Yang-Mills theory and
an exactly marginal (in 2d sense)
deformation of \adss superstring theory
was suggested in \ci{LM} and further explored in \ci{frt,f}.

Here we shall be interested in generalizing
the results  of \ci{frt}  about the correspondence between
semiclassical string states  and ``long''  gauge-theory operators
to the case of 3-spin $(J_1,J_2,J_3)$
string states dual to operators built out of the three
holomorphic combinations
of 6 real scalars (analog of $SU(3)$ in undeformed theory).
The comparison between  string and gauge theory
in this  sector turns out to be quite non-trivial.

We shall consider the case of
real deformation parameter $\b\equiv \g - i \bfsigma  = \g$.
It turns out to be straightforward  to  generalize the discussion to
 the case of the more general non-supersymmetric
 3-parameter ($\g_i$)  deformation of the \adss geometry
% (preserving $[U(1)]^3$ subgroup of $SO(6)$)
constructed in \ci{f}
using the same TsT (T-duality, shift, T-duality) transformation as in \ci{LM}.
This deformation is quite  natural as it treats all 3 isometric angles
 of $S^5$ on an equal footing.
The corresponding type IIB  supergravity background preserves
1/4 of supersymmetries (8 supercharges) only
in the  ``symmetric''   LM  \ci{LM} case
\be \la {lm}  \g_1=\g_2=\g_3 =\g \ . \ee
However, as we will see, this symmetric point is not special as far
as the correspondence between string and gauge theory is concerned:
 the matching of leading-order semiclassical  string energies and
 one-loop
gauge theory anomalous dimensions we are  going to establish below
holds in the general $\g_i$ case.

%v2
This appears to be  one of the first non-trivial  examples
 when  implications of the
AdS/CFT  duality are  observed  at a quantitative level
in a  {\it non}-supersymmetric  
case.\foot{The present case is obviously different from 
the   examples of 
(non)supersymmetric orbifolds \ci{kas} of the  \adss -- $\N=4$ SYM 
duality 
where large $N$ duality relations are ``inherited'' in untwisted 
sector. 
Same applies to  the type 0  
 analog  of the AdS/CFT 
  duality \ci{KT}  obtained by $(-1)^F$-type  orbifolding;
 a discussion of  matching of some of string energies and gauge theory 
 anomalous dimensions
in the BMN limit of type 0 theory appeared in \ci{cot}.}
It  provides a strong motivation for
further study of this $\g_i$-dependent string theory and the
conjectured dual
non-supersymmetric large $N$ gauge theory \ci{f}
is of obvious interest and importance. One particularly   interesting
aspect
is the existence (for certain range of parameters)
of closed-string tachyons  and their  reflection on the gauge theory side.
This non-supersymmetric  theory is certainly stable  in the
nearly-flat  and small $\g_i$  limit  and thus appears to be more
under theoretical
control than the type 0 example  considered in \ci{KT}.

\bigskip

We shall start in section 2  with presenting the 3-parameter deformation
of the \adss background found  by the direct  generalization of the LM
construction in \ci{f}.
We shall then discuss  the BPS states  and  more  general
 geodesics on $\g_i$-deformed $S^5$ representing semiclassical
 point-like string
 states.
The  geodesics happen to be described  by a 1-d
 integrable  Neumann  model
which is the same as  the  system describing 
rotating \ci{ft2} and pulsating \ci{minah}
circular strings in
$S^5$ part of \adss  \ci{art}. The solutions  are labeled in general
by 3 conserved angular momenta $(J_1,J_2,J_3)$  and one  additional
integral of motion  (``oscillation number'') and
 depend on deformation parameters $\g_i$ through the combinations
\be \la{nuu}
\nu_i\equiv \ep_{ijk} \g_j J_k \ .   \ee
%SF
These combinations are the twists that appear in the relations between
the angle variables
of $S^5$ and the $\g_i$-deformed five-sphere \ci{f}.
By using these  relations
%By following the TsT transformation that relates  \ci{f} the
%\adss string theory  to  the $\g_i$-deformed string theory
one can show that  in the special cases  when
  $\nu_i$ are   integer the
circular pulsating and rotating
 strings of undeformed theory are, indeed,  the
images of the point-like strings in the deformed
 geometry, with $\nu_i$  being the counterparts  of the
 circular string winding numbers $m_i$.\foot{If $\nu_i$ are not integer the formal
images  of geodesics of deformed geometry in \adss theory
  do not satisfy closed-string periodicity conditions.
%SF
These images are open strings subject to twisted boundary conditions.
  }
While in the standard \adss (undeformed)
 case all geodesics were  representing BPS states
with energy $E$ equal to the total angular momentum $\JJ=J_1+ J_2 + J_3$
here we shall find that only a few of them
 have this ``vacuum state''   property. These special ``BPS'' geodesics
 have energies that do not depend
on the deformation parameters, i.e.
are the same as in the undeformed case. They can be labeled
by the angular momenta as:
(i) $(\JJ,0,0),\ (0,\JJ,0),\ (0,0,\JJ)$ and (ii) $(J_1,J_2,J_3)_{vac}$ with
$ \nu_i=0$, i.e.
\be \la{vac}
J_{i, vac} = {\g_i \ov \gga} \JJ\ , \ \ \ \ \ \ \ \ \ \ \ \
 \gga\equiv  \g_1 + \g_2 + \g_3 \ . \ee
The $\nu_i=0$  condition  is
 satisfied for the $(J,J,J)$  BPS state \ci{LM}
 in the symmetric LM case of $\g_i= \g$. In general,
since $J_i$ should take integer values in  quantum theory,
such states will exist only for special  choices of $\g_i$.
In addition to these special BPS states  which are images of the corresponding
point-like ($\nu_i=m_i=0$)
or BPS states  of the undeformed theory,
there is another simple subclass of geodesics
for which radial directions are constant in time:
these  are (for integer $\nu_i$) the
 TsT images of
rigid rotating circular strings \ci{ft2,art} in undeformed $S^5$.
Their classical energy has non-trivial dependence on $J_i$ and $\g_i$
and receives also string $\a'$ corrections.

As in the undeformed case,  it is straightforward  to explore the
fluctuation spectrum  \ci{BMN}
 near  particular geodesics, i.e. quantum energies
of semiclassical
``small'' (nearly point-like)  string states in the limit of large
total angular momentum $\JJ$. The spectrum near the $(J,0,0)$
geodesic is similar to the standard BMN one \ci{NIPR,LM}.
In the case  of the
expansion near the $J_i \sim \g_i$ geodesic \rf{vac}  the
 spectrum of small $\s$-dependent fluctuations
turns out to be
independent of the deformation parameters, i.e. to be the same  as
the BMN spectrum in the undeformed theory.
 The same conclusion  was reached earlier in the symmetric LM $\g_i=\g$ case
in \ci{mat,koch}. This, in fact, is implied 
(to leading order in $1/\JJ$)
by the
 TsT transformation  of  \ci{f}.
%We  shall show in section 6 that the same
%result is found on the gauge theory  side.
We shall discuss the spectrum of fluctuations 
on the gauge-theory side in  Appendix A.
The zero-mode part of the spectrum (corresponding 
to fluctuations
depending only on
time,  i.e.
within the space of  geodesics of deformed theory)
  is, however,
non-trivial \ci{mat}; 
%
%RR
%
%%we shall discuss its relation to gauge theory 
%%in \ref{app:zeromode} where
%
we shall  match it  with the one-loop
gauge theory  prediction in Appendix B.

\bigskip

In section 3,  we shall turn to other  semiclassical  states
represented by  extended  strings moving  fast  in deformed $S^5$.
As in \ci{frt},  they   can be systematically described by  reducing the
 classical string
action to a kind of ``Landau-Lifshitz'' (LL)  sigma model \ci{KRUC,kt}
for the ``transverse'' string degrees of freedom.
In the present 3-spin case  we  shall obtain   a deformed version
of the $\IC\IP^2$  LL model  corresponding  to  the $su(3)$ sector of
 the \adss string  theory
\ci{hlo,st,kt}. As in the deformed 2-spin  case of \ci{frt}, we shall find
 that the
deformed 3-spin LL model  contains a potential
 term which is responsible for lifting the
energies of all of the string states apart from  few
 BPS  ones (the point-like states discussed above  and some circular
 BPS strings existing as in \ci{LM}  for special $\g_i$).

The  challenge  will then  be to  find the counterpart of this  action
 on the gauge-theory side  and to show that it coincides with
 the string expression; this would imply, in particular, the agreement
between
 the leading correction to string energies and
 one-loop anomalous dimensions of the corresponding gauge-theory
 operators.

\bigskip

 In section 4 we  shall
  present  the direct  generalization \ci{RR} (see also \ci{BECH})
  of the 1-loop dilatation operator
 for the   exactly marginal $\b$-deformation
 \ci{LEST,old}  of $\N=4$ SYM  to the
 non-supersymmetric case of the three $\g_i$ deformation parameters.
 As in the symmetric $\g_i=\g$ case, it
 can be identified with  an integrable  spin chain Hamiltonian
 (with 3 spin projections at each site corresponding to 3 chiral  scalars $\P_i$)
 which is a deformation of the $su(3)$ invariant  XXX$_1$  Hamiltonian
 \ci{mz}.
 We shall then   describe   the corresponding
 generalization of the Bethe ansatz equations and 
   apply them  to show
  that the  ground states
 of the 1-loop spin chain Hamiltonian  are indeed the same as found
 on the string side.
 %AT
 We shall also discuss the distinction  between 
 the $U(N)$ and $SU(N)$ gauge group cases which survives here the large $N$
 limit since the $U(1)$ parts of matter  fields do not decouple. 
 
 \bigskip

 In section 5 we shall finally turn to the derivation of the
 effective  coherent-state action for low-energy  semiclassical
 states  of the spin chain that should be dual to the semiclassical
string states in the 3-spin sector.
In general, there are many equivalent
spin-chain Hamiltonians, corresponding to  different choices of basis
in the space of gauge-theory operators,
that lead to the same  anomalous dimensions.
To establish the
correspondence with string theory it turns  out  that one needs
 a special choice adapted to low-energy approximation.
This is a  subtlety  not confronted in previous discussions of the coherent state
approach in the undeformed  \ci{kt}
or deformed 2-spin  \ci{frt} cases.
We shall  describe   the choice of coherent states and the basis needed
to capture the expected   BPS  states \rf{vac} in low-energy
 (slowly-changing coherent field)
approximation in sections 5.1 and 5.2.
  Then in section 5.3 we shall find
  that this  choice leads exactly to the same
  Landau-Lifshitz  effective action as found  in section 3 on
the string side.
This   provides a highly nontrivial  check of
the AdS/CFT duality not only in  the supersymmetric LM deformation case \ci{LM}
but  also in the general
 non-supersymmetric $\g_i$-deformed theory.

\bigskip

Section 6 will contain some concluding remarks.

In Appendix A   we shall discuss  fluctuations  near 
the vacuum
states of the one-loop spin chain
and  match  their spectra  with the string-theory
 results. 
%
% RR
%
In Appendix B we shall consider 
 the   spin-chain
 0-mode fluctuations  near the $(J_1,J_2,J_3)$ vacuum
 and again demonstrate remarkable agreement with the string-theory 
 predictions.
%  from
%section 5 will be presented in Appendix A.

\bigskip

%%%%%%%%%%%%%%%%%%%%%%%%%%%%%%%%%%%%%%%%%%%%%%%%%%%%%%%%%%%%%%%%%%%%%%%%%%%%%%%%%%%%

\renewcommand{\theequation}{2.\arabic{equation}}
 \setcounter{equation}{0}

%%%%%%%%%%%%%%%%%%%%%%%%%%%%%%%%%
\section{Three-parameter  deformation of \adss string theory}
%%%%%%%%%%%%%%%%%%%%%%%%%%%%

%%%%%%%%%%%%%%%%%%%%%%%%%%%%%%%%%%%
\subsection{Background}
%%%%%%%%%%%%%%%%%%%%%%%%%%%%%%%%%%%%%

We shall mostly follow the notation of \ci{frt}.
The type IIB solution  related by T-dualities and shifts   transformation
to the \adss  background  and which generalizes \ci{f} the background of
\ci{LM} to the case of unequal $\g_i$ parameters can be represented  as
\begin{eqnarray}
ds^2_{\rm str} &=& R^2  \left[ ds^2_{_{AdS_5}} +
   \sum^3_{i=1} ( d\rho_i^2  + G \rho_i^2 d\phi_i^2) +   G
\rho_1^2\rho_2^2\rho_3^2 [d (\sum^3_{i=1} \tg_i \phi_i)]^2 \right]
\label{me}
\\
\la{bfield0}
B_2 &=&  R^2  G w_2 \ , \ \ \ \ \ \ \
 w_2  \equiv  \tg_3  \rho_1^2 \rho_2^2 d\phi_1 d\phi_2 +
\tg_1  \rho_2^2 \rho_3^2 d\phi_2 d\phi_3 + \tg_2
\rho_3^2 \rho_1^2 d\phi_3 d\phi_1\ ,   \\
e^\phi &=& e^{\phi_0}G^{1/2} \, ,\ \ \ \ \ \    \qquad\chi = 0 \, ,
\\
\label{ofggen0}
G^{-1} &\equiv& 1 +  \tg_3^2  \rho_1^2 \rho_2^2 +  \tg_1^2
\rho_2^2 \rho_3^2 +  \tg_2^2  \rho_1^2 \rho_3^2
~,\qquad \ \ \ \ \sum_{i=1}^3\rho_i^2=1 \ ,
%R_E^4 = 4 \pi N
\\
 C_2  &=& -  4R^2 e^{-\phi_0}    w_1 d ( \sum^3_{i=1} \tg_i \p_i)
\ , \ \ \ \ \ \ \ \ \ \ \     d w_1 \equiv     \cos \alpha
\sin^3\alpha \sin {\theta}  \cos {\theta}
d\alpha d\theta ~,
\\
F_5 &=& 4 R^4 e^{-\phi_0}  ( \omega_{_{AdS_5}} +  G \omega_{_{S^5}})
~,\qquad\omega_{_{S^5}} \equiv { }  dw_1 d\phi_1 d\phi_2 d\phi_3
\end{eqnarray}
Here $B_2$ is the NS-NS 2-form potential, $\phi$ is the dilaton
and $d\chi, \ dC_2 $ and $F_5$ are the R-R field strengths.
 The angles  $\theta$,$\alpha$  appearing in $dw_1$
parametrize $S^2$  coordinates $\r_i$  as follows
\be\la{angl}
\rho_1 = \sin \a \cos \th\ , \ \ \
\r_2 =  \sin \a \sin \th\ , \ \ \
 \r_3=  \cos \a\
 .
\ee
Note also that
\bea\la{w1}
w_1= {1\over 4}\,\r_1^2\,d(\r_2^2)-{1\over 8}\,d(\r_1^2\r_2^2)\=
{1\over 8}\,(\r_1^2+\r_2^2)^2 d {\r_2^2\over \r_1^2+\r_2^2}
\ .
\eea
% (as in \cite{LM} we use  short-hand notation like
%$c_\alpha\equiv  \cos \alpha$ and  $s_\alpha\equiv  \sin \alpha$.).
The standard \adss background is recovered  after setting the deformation
parameters $\tg_i =R^2 \g_i  $  to zero.
For equal $\tg_i=\tg$ this becomes the background of \ci{LM}
($\tg_i$ were denoted as $\hat \g_i$ in \ci{LM,f}).
We also assume that
\bea
g_s &=&  e^{\phi_0}  =
{g^2_{\YM}\ov 4\pi}\, , \ \ \ \ \
\quad R^4= 4 \pi g_s N =N
g^2_{\YM} \equiv\lambda \ , \ \ \ \ \ \ \ \ \   \alpha'=1\ , \la{p}   \\
\tg_i &=&   R^2 \g_i = \sql \g_i  \  .\la{fgi}   \eea
Here $\g_i$ are the deformation parameters which
appear on the gauge theory or spin chain side.
  In the
symmetric case $\g_i=\g$ this parameter is the real part of the deformation
parameter $\b$ in the superpotential
$W= \, h\ \Tr( e^{ i  \pi \beta} \Phi_1\Phi_2 \Phi_3 - e^{- i  \pi \beta}
 \Phi_1\Phi_3 \Phi_2  )$.
  We shall  consider only the case of real $\b$
where the  duality appears to be much more under quantitative control
(see \ci{frt}).

As discussed in \ci{frt}, the parameters $\tg_i$  which enter  the
supergravity background are
assumed to be fixed in the semiclassical string limit.
Since ${R^2\ov \a'} = \sql$ plays the role of the string tension,
in this limit one also fixes  other semiclassical parameters like
$\E$ and $\J_i$  which determine the string energy and  spins
\be
E= \sql \E \ , \ \ \ \ \ \
J_i = \sql \J_i \ , \ \ \ \ \
\tl \equiv {\l \ov \JJ^2}={\rm fixed} \ ,\ \ \ \ \  \ \ \JJ=\sum^3_{i=1} J_i \ ,
\ee
while  $\sql$ and thus $\JJ$ are assumed to be large to suppress string
$\a'$ corrections.  That means that
\be
\bg_i \equiv \g_i \JJ = {  \tg_i \ov \sqrt{\tl} }
\ee
is also fixed in this limit, i.e. $\g_i \sim { 1 \ov \JJ}$.
%NEW
For definiteness, we shall assume that both $J_i$ and $\g_i$ are non-negative.

On the gauge theory (spin chain)  side,  the limit which one takes
is formally different \ci{BDS,frt}.
Since one uses  perturbative gauge theory, one first expands in
$\l$ and then takes  $\JJ$ large.
Here $\JJ$  plays the role
of the length of the chain (or length of the operator),
and we will be interested in extracting the dependence of the
spin chain energies  on the
parameters $\tl$ and $ \g_i \JJ$  while
looking at 1-loop (order $\l$) correction and taking large $\JJ$ limit.
In all previously discussed examples of similar comparisons
the leading order terms in the two  expressions matched,
and our aim will be to extend this matching to the
present (non-supersymmetric  for unequal $\g_i$) case.

%SF
%%%%%%%%%%%%%%%%%%%%%%%%%%%%%%%%%%%
\subsection{BPS states}
%%%%%%%%%%%%%%%%%%%%%%%%%%%%%%%%%%%%%
By following the TsT transformation
that relates  the $AdS_5\times S^5$ string theory to the
$\g_i$-deformed string theory one can relate
 the angle variables $\ttp{}_i$ of $S^5$ (in the notation of \ci{f})
 and  the angle variables $\p_i$ of the TsT-deformed geometry \rf{me}.
The basic starting point  is the
equality  between the $U(1)$ conserved current densities  of strings on
$AdS_5 \times S^5$ and on the $\g_i$-deformed background \ci{f}:
\bea
\widetilde{\widetilde{{\bf J}}}{}_{i\ p} = {\bf J}_{i\ p} \ ,
\label{rel1}
\eea
where $i=1,2,3$ and $p=0,1$ are the world-sheet indices.
Taking into account that the time components of the currents
 are the momentum densities  conjugate to the angle variables,
and expressing the time derivatives through the momenta,
one can cast   (\ref{rel1}) in the
following simple form
\bea
\widetilde{\!\widetilde{p}}\,{}_i &=& p_i = {\bf J}_{i\ 0}\ ,
\label{rel2}\\
\r_i^2\,\widetilde{\!\widetilde{\p}}\,{}_i' &=& \r_i^2 ( \p_i' - \epsilon_{ijk}\g_j p_k )\ , \ \ \
\ \  i=1,2,3 \ ,
%\quad{\rm no\ summation\ over\ } i\ .
\label{rel3}
\eea
where in \rf{rel3}  we assume summation in $j,k$ but no summation in $i$.
If none of the ``radii'' $\r_i$ vanish on a string solution,
one can cancel the $\r_i^2$ factors in (\ref{rel3}) to  get
\bea
\widetilde{\!\widetilde{\p}}\,{}_i' = \p_i' - \epsilon_{ijk}\g_j p_k\ .
\label{rel4}
\eea
Integrating over $\s$ and taking into account that $\p_i$ are angle
variables
and the strings in the deformed background are assumed to be
closed, i.e.
\be \p_i(2\pi)-\p_i(0)= 2\pi n_i \ , \ee
where $n_i$ are integer winding numbers,
 we get the twisted boundary
conditions for the angle variables
$\,\widetilde{\!\widetilde{\p}}{}_i$
of the original $S^5$ space
\bea
%\p_i(2\pi)-\p_i(0)&=&2\pi n_i\ , \quad n_i\ {\rm are\ integer\ winding\ numbers}\\
&&
\widetilde{\!\widetilde{\p}}_i(2\pi)
-\widetilde{\!\widetilde{\p}}_i(0)=2\pi(n_i - \nu_i)\ , \
\label{rel5}\\
&& \nu_i \equiv  \epsilon_{ijk}\g_j J_k\ , \ \ \ \ \ \ \ \ \ \ \
J_i = \int^{2 \pi}_0 { d \s \ov 2 \pi} \  p_i  \ . \la{nui}
\eea
% $\nu_i$ are the twists already mentioned  in (\ref{nuu}).
We see that if the twists $\nu_i$ (already mentioned in  (\ref{nuu}))
 are not integer then
 the twisted strings in \adss  which are formal images of closed strings in the deformed
 geometry  under the inverse of TsT transformation  are open.

The relations (\ref{rel3}) imply that if $\p_i$ solve the
equations of motion for a string
in the $\g_i$-deformed background then
$\,{\widetilde{\!\widetilde{\p}}}{}_i$ solve those
 in \adss with
the twisted boundary conditions (\ref{rel5}) imposed on the angle variables. It is easy to show \ci{f}
that the Virasoro constraints for both models also
map to each other under the TsT-transformation;
therefore, the energy of  a twisted string in \adss is equal to the energy of the
corresponding closed string in the $\g_i$-deformed background.
 This observation allows one to readily
determine  all classically BPS 
%%(minimal energy for given charges)
states in the deformed model, i.e. the states that have
 minimal energy for the given
charges,
\be \la{enr}
E=\JJ\equiv   J_1 + J_2 + J_3  \ . \ee
To this end we notice that a BPS state in the deformed background must be
an image of a BPS state
in \adss, that is an image of a point-like string or null 
geodesic in \adss.
For such a string $\widetilde{\!\widetilde{\p}}\,{}_i{}' = 0$, \ 
$\r'_i=0$; then
 $\widetilde{\!\widetilde{p}}\,{}_i = p_i = J_i$ do not depend on $\s$,
i.e.  all the charges
are distributed uniformly along the string. Thus, for the BPS states
the relation (\ref{rel4}) takes the form
\bea
\p_i' = \epsilon_{ijk}\g_j p_k = \nu_i\ ,\ \ \ \ \ \ \ \ \  p_i = J_i \ ,
\label{rel6}
\eea
where we also assume that all the  charges $J_i$ are not equal to 0.
Since the string in the deformed background is closed,
 all the twists $\nu_i$  which play the role of the winding numbers
 then must be integer:
\bea
\label{rel7}
\nu_i = \epsilon_{ijk}\g_j J_k\ \ \in\IZ~~.
\eea
One is now  to distinguish the case of non-zero $\nu_i$
when a solution is a circular string, and the case of $\nu_i=0$
when the solution is a point-like string.

For $\nu_i\neq 0$ these equations can have a consistent
(circular) string solution   only
if $\g_i$ are rational ($J_i$  take integer values in quantum theory)
and the corresponding BPS state
is a circular string similar to the ones studied in
\ci{ft2}
(this generalizes the observation in \ci{LM} to the
 case of unequal $\g_i$).

For $\nu_i=0$ the BPS state   of deformed  geometry is a point-like string.
 The general
solution to $\nu_i=0$ is
\bea
\nu_i=0 \ : \ \ \ \ \ \ \ \  \ \ \  J_i = c \g_i \ ,
%\g_i = \underline{\g} m_i\
%; J_i= \underline{J} m_i
\la{rel8}
\eea
% $\underline{\g}$ is any real number, $m_i$ and $\underline{J}$ are
% any integers.
where $c$ is a proportionality coefficient which can be any real number.
Since $J_i$ must be integer in the quantum theory,  such a
solution exists only for special
values of $\g_i$.\foot{It is interesting that since for the classical
strings there is no quantization condition, any solution
satisfying
$J_i\sim \g_i$
has the BPS energy $E = J_1 + J_2 + J_3$. It would be interesting to
analyze
the semi-classical expansion around such a solution,
and to see how the quantization condition gets restored.}
These $(J_1,J_2,J_3)$ point-like
BPS states generalize the $(J,J,J)$ state \ci{LM} in the supersymmetric
LM  case $\g_i=\g$, $J_i = J$.

Note that   any  $(J_1,J_2,J_3)$
   solution  in the deformed background
   for which  \rf{rel8} is satisfied
   can be obtained from a closed string solution in \adss,
   and  the energies of these  string states in the $\g_i$-deformed model and
   their images in
   the  \adss are equal to each other.\foot{Such (in general, 
   non-BPS) states
%One
%can may say that
%Such  string states may
should  be  dual to the gauge-theory  operators 
protected from the deformation
at least to the leading order in $\g_i$.}

If one of the 3 momenta is equal to zero, e.g., $J_3=0$,
then the string states belong to the 2-spin sector
which is the analog of the  $su(2)$ sector of undeformed theory.
%In this  case we find also
%BPS state is $(J,0,0)$ for any choice of $\g_i$.
It contains  the obvious additional  BPS state $(J,0,0)$
which is the direct TsT relative of the corresponding point-like state in \adss.
Similarly, we have also $(0,J,0)$ and $(0,0,J)$  BPS states.

%%%%%%%%%%%%%%%%%%%%%%%%%%%%%%%%%%%
\subsection{Point-like strings (geodesics) and near-by fluctuations}
%%%%%%%%%%%%%%%%%%%%%%%%%%%%%%%%%%%%%

Let us now  analyze some
string solutions  in the deformed geometry starting directly
with \rf{me},\rf{bfield0}.

 To find  the classical point-like string states
in the  deformed  geometry
it is enough to concentrate on the string-frame metric
(to study quantum corrections one will need of course the full
Green-Schwarz fermionic action which will contain couplings to other
 background fields). We should consider
 geodesics that wrap the ``internal'' $S^5_\g$  part
and that  should be dual to special
$\Tr ( \P_1^{J_1} \P_2^{J_2}\P_3^{J_3}+...)$
operators on the gauge theory side.

 The metric \rf{me} has 3 isometries corresponding to
shifts of the angles $\p_i$ and thus  the states should be characterized by
3 conserved angular momenta $J_i$. Starting with the string equations in
conformal gauge
with $AdS_5$ time $t= \E \tau$  it is straightforward to show that,
while the metric looks rather complicated,
the effective  action that determines the  time evolution
of the $S^2$ coordinates $\r_i$  can be written simply as
($\sum^3_{i=1} \r^2_i=1$)
\be  \la{par}
S(\r)= \ha \sql \int d \tau \  L\ , \ \ \ \
L(\r) =  \sum^3_{i=1} \big[  \dot \r_i^2 - V_i (\r_i) \big] \ ,
\ \ \ \ \ \ \ \ \ \ \
V_i (\r_i) = { \J^2_i \ov \r^2_i}  +  \nu^2_i \r^2_i \ , \ee
\be\la{defo}
J_i = { \del L(\r,\p) \ov \del \dot \p_i}= \sql \J_i \ ,
\ \ \ \ \  \ \ \ \ \
   \nu_i \equiv  \ep_{ijk} \tg_j \J_k  = \ep_{ijk} \g_j J_k   \ . \ee
 Here   $L(\r,\p)$ stands for the string Lagrangian before one
 solves for the derivatives of the angles.
For $\nu_i=0$ this is  the  action of a particle moving on $S^5$.
For general $\nu_i$ this is recognized as a Neumann-Rosochatius
 integrable system
describing an oscillator on 2-sphere (or, equivalently, a special Neumann
system describing an oscillator on 5-sphere, cf. \ci{art}).
The conformal gauge constraint
 implies  that the corresponding Hamiltonian is equal to
$\E^2$, i.e.  $\sum^3_{i=1} \big[  \dot \r_i^2 + V_i (\r_i) \big] = \E^2$.
In particular,  in the LM case of $\tg_i=\tg$ we get explicitly for the
particle Hamiltonian
\bea
H=\E^2  =   \dot \r_1^2 + \dot \r_2^2 + \dot \r_3^2
   &+&
{ \J^2_1 \ov \r^2_1}  +   { \J^2_2 \ov \r^2_2}  + { \J^2_3 \ov \r^2_3}  \nonumber \\
&+&     \tg^2  \left[  (\J_2 - \J_3)^2  \r^2_1
+  (\J_3 - \J_1)^2  \r^2_2 + (\J_1 - \J_2)^2  \r^2_3\right]
  \ . \la {jo} \eea
This result is easy to find using the TsT relation \ci{f}
of the deformed theory to the \adss theory. The two string
Hamiltonians are related by  the TsT,
so to get the particle Hamiltonian in the deformed theory
all  one has
 to do is to shift  the $\s$-derivatives of the  \adss angles
%$ \widetilde{\!\widetilde{\p}}\,{}_i'$
 by the momenta as in \rf{rel4}
and then to set all terms with  $\s$-derivatives to zero.

The appearance  of the Neumann system
is not accidental: the same system was found in \ci{art}
to describe  circular pulsating
and rotating strings in undeformed $S^5$. These strings
are, in fact,  mapped (for integer $\nu_i$)
 to
point-like strings in $S^5_\g$   under the  TsT transformation of \ci{f}:
$\nu_i$ plays the role of the winding number $m_i$ of the circular strings, and
the conformal gauge constraint $m_i J_i=0$ here is satisfied automatically.

Generic solution is labeled  by $(J_1,J_2,J_3)$ and one extra
  (in addition to the 1-d energy or $H=\E^2$) integral of motion
which may be interpreted as an ``oscillation number'' $K$ for a
(quasiperiodic)
particle motion  on $S^2$. The lower-energy solutions
correspond to $K=0$ when $\r_i=$const. The form of the dependence of the energy
on $K$ and $J_i$ will be the same as in the case of the pulsating strings in
\ci{Mina,kt}.

The special solutions  that are the same as in the undeformed case
and thus represent the  lowest-energy (``BPS'')  states
are found if (i) $\r_1=1, \ \r_2=\r_3=0$ (and two other cases with  interchange
of 1,2,3), representing  $(J_1,0,0)$ state with  $E= J_1$, and  also 
if (ii)
$J_i$ are such that $\nu_i=0$, i.e. if $J_i\sim \tg_i$
when
$
E= \JJ= J_1 + J_2 + J_3 $.
These are the same as already discussed in the previous subsection and
they  should be dual to vacuum (zero anomalous dimension) states
on the spin chain side.

In the general non-supersymmetric case of  unequal $\g_i$
there is an open question if such states are  true vacua
%NEW
(i.e. states with $E=\JJ$ which is the  absolute minimum of the energy),
i.e. do not
 receive quantum corrections both on the string theory and on the gauge
 theory side.\foot{
It is not a priori clear  that string  $\a'$ corrections are  absent:
while TsT  transformation does not affect these special geodesics,
it may (and, in fact, does) change the  spectrum of fluctuations near them,
and thus may alter the cancellation of the quantum correction to the
 vacuum energy.
% In general, one can of course study the expansion about any local minimum;
% the local minimum  receives quantum corrections but
 % the perturbative theory is
%well-defined.
}

 % {\bf this is a calculation that needs careful fixing
%of fermionic part -- can be done; done by Mateos in symmetric case.}

In addition, there are higher energy (non-BPS) states
still having $\r_i=$const, i.e. $K=0$, which (for integer $\nu_i$)
are images of rigid (non-pulsating)
circular rotating strings in $S^5$.
As discussed in \ci{ft2,art}, the
classical energy
of the latter is a non-trivial  function  of $J_i$, the winding numbers $m_i$ and
the string tension; expanded
 in $\tl$ it   looks like
 $E= \JJ + {\l\ov \JJ} c_1(m_i, {J_j\ov J_k}) + ...$.
The same expression is   found  for the point-like strings
here  with $m_i \to \nu_i$.
The leading order corrections to $E=\JJ$ relation will  scale as
${\l\ov \JJ} (\g_i J_n)^2 ({J_j\ov J_k})^2 $ and may
thus  be compared to the gauge-theory side.
We will do this automatically by matching the corresponding effective actions
that describe such semiclassical states.

\bigskip

Next, let us  follow \ci{BMN,GKP2} and  study small
semiclassical strings representing small fluctuations near the
above geodesics.

%%%%%%%%%%%%%%%%%%%%%%%%%%%%%%%%%%%%%%%%%%%%%%%%%%%
\subsubsection{ $(\JJ,0,0)$ case}
%%%%%%%%%%%%%%%%%%%%%%%%%%%%%%%%%%%%%%%%%%%%%%%%%%%%%%%

In the $\g_i=\g$ case  the corresponding analog of the BMN spectrum
of quadratic fluctuations near the $(\JJ,0,0)$  geodesic
 was found in \ci{NIPR,LM}. Here we shall generalize it to the
non-supersymmetric $\g_i$-case.
We shall first concentrate on the bosonic part of the fluctuation Lagrangian that
follows {}from expanding the  bosonic part of the string action
which depends only on the string metric \rf{me}
and the 2-form field $B_2$ in \rf{bfield0}
\bea
\la{A}
I_B = -\ha {\sql}\int\, d\tau\int^{2 \pi}_0   {d\s\over
2\pi} \left[ \sqrt {-g } \ g^{pq}\partial_p X^M\partial_q X^N\, G_{MN}
-
\e^{pq}\partial_p X^M\partial_q X^N\, B_{MN}\right]\, ,
\eea
where  $\e^{01}=1$ and in the conformal gauge which we shall use here
$g_{pq} = \mbox{diag}(-1,1)$.
Expanding the action near the solution $t=\phi_1= \J \tau,
 \ \r_1=1, \ \p_{2,3}=\r_{2,3}=0$
we get for the part of the  fluctuation Lagrangian
which is different from the standard BMN $\g_i=0$ case
\bea
L = \ha (\dot y_a^2 - y_a'^2  + \dot z_a^2 - z_a'^2 )
 &+&\ha \J^2 (1+\tg_3^2) y_a^2  + \ha \J^2 (1+\tg_2^2)  z_a^2 \nonumber \\
 &+&  \J \tg_3 \ep_{ab} y_a y_b'
   + \J \tg_2 \ep_{ab} z_a z_b'
%\ \ \ \ \   \J= \J_1
 \ .  \la{joo} \eea
Here we assume summation over $a,b=1,2$  and
 $y_a$ and $z_a$ are 2+2  fluctuations of Cartesian coordinates
in the $\r_2,\p_2$ and $\r_3,\p_3$ planes.
This is essentially the same Lagrangian as in the $\g_i=\g$ case \ci{LM,NIPR}
but with the parameters $\g_2$ and $\g_3$  in the each of the 2-planes
transverse  to the geodesic. The expansion near $(0,\JJ,0)$ and $(0,0,\JJ)$ geodesics
leads to similar expressions with the corresponding interchange of the
parameters $\g_i$.

 The corresponding characteristic frequencies
that represent  the analog of the BMN spectrum $E-\JJ = 
 { w_n \ov \J} N_n $ are
($n=0,\pm 1, ...$ labels string $e^{i n \s}$ modes and is different for 
the two types for the excitations)
\be \la{fre}
w^{(i)}_n =  \sqrt{\J^2 + ( n + \tg_i \J)^2} =
\J \sqrt{ 1 + \tl (n + \g_i \JJ)^2 } \ ,
\ \ \ \ \   i=2,3 \ . \ee
As follows from the structure of the  supergravity background,
the quadratic fermionic action contains couplings  only to the
NS-NS 3-form (with two parts proportional to $\g_2$ and $\g_3$
as reflected in \rf{joo})  and the standard R-R
5-form flux. It thus  has the structure as  in eq.(4.12) in \ci{NIPR}
with $\g_2$ and $\g_3$  multiplying the corresponding fermionic projectors
$\theta^A\Gamma^+ ( \tg_2 \G_{y_1y_2} + \tg_3 \G_{z_1 z_2}) \theta^A$
and a mass term coming from 5-form flux
 (see also \ci{mets} for a general structure of such actions).
Then (as  follows, e.g., from eq.(4.21) in  \ci{NIPR})
the corresponding fermionic spectrum is the same as the above bosonic one,
implying that the quadratic fluctuation Lagrangian  has 2-d
world-sheet supersymmetry.
The latter  is a consequence of space-time
supersymmetry of the corresponding plane-wave
background  (for which \rf{joo} is the l.c. gauge fixed Lagrangian)
present even though the original supergravity background is not supersymmetric
for unequal $\g_i$.
This has an important consequence that the contribution of the  quadratic
fluctuation energies to the $(\JJ,0,0)$  ground state energy vanishes,
i.e. (at least to the leading order in $1/\JJ$)
 this state  is a  true analog of the corresponding  BPS  state in
 the undeformed or in the supersymmetric deformed
 $\g_i=\g$ theory.

We shall see that these conclusions  are corroborated by the
 analysis of the one-loop dilatation operator on 
 the gauge theory side.
In particular, the same  fluctuation spectrum
(for the relevant part of fluctuations)
will appear from the  coherent
state action for  the 3-spin or 
the holomorphic 3-scalar  sector
of the spin chain  which is the analog of $su(3)$ sector in
the undeformed theory.

%%%%%%%%%%%%%%%%%%%%%%%%%%%%%%%%%%%%%%%%%%%%%%%%%%%
\subsubsection{ $(J_1,J_2,J_3)_{vac} $  case}
%%%%%%%%%%%%%%%%%%%%%%%%%%%%%%%%%%%%%%%%%%%%%%%%%%%%%%%

While the $(\JJ,0,0)$ case is very similar to the standard BMN case, the
expansion near the
$(J_1,J_2,J_3) \sim (\g_1,\g_2,\g_3)$ geodesic
is more involved.
The bosonic part of the fluctuation Lagrangian
follows from \rf{A}  expanded near
the corresponding classical solution (again, we assume that $\g_i \geq  0$)
\be  \la{sol}
t= \p_i = \J \tau  \ , \ \ \ \ \ \ \ \ \
\r^2_i = {\J_i \ov \J}= {\g_i\ov\gga}   \ , \ \ \ \ \ \ \
\J \equiv  \sum^3_{i=1} \J_i \ ,\ \  \ \ \gga \equiv   \sum^3_{i=1} \g_i    \ . \ee
The fluctuations
in time and $\psi = \sum^3_{i=1} \tg_i\p_i$  directions decouple, i.e.
are massless 2d fields,
 the fluctuations in the other 4 $AdS_5$  directions are the same as in the
undeformed case (i.e. are described by massive 2d fields with mass
$\J = {\JJ\ov \sql}$)  while the remaining 4 non-trivial
fluctuations in $S^5_\g$ directions  are  found by setting ($a=1,2$)
\be \la{io}
\p_a = \J \tau + v_a \ , \ \ \ \   v_3= - { \g_1\ov \g_3} v_1 - { \g_2\ov \g_3} v_2
\ , \ \ \
\r_a = \sqrt{\g_a \ov \gga} ( 1 +  u_a)\ , \ \ \ \ \
\r_3=\sqrt{1- \r_1^2 -\r_2^2}  \ , \ee
where $v_1,v_2,u_1,u_2$ are 4 independent 2d  fluctuation  fields.
Computing the   momenta  for $\p_i$, i.e. the angular momenta $J_i$, 
from the Lagrangian \rf{A} we get, to the leading order in  
fluctuations near the vacuum \rf{io}: 
\be \la{jes}
J_i  = {\g_i\ov\gga}  J + j_i  \ , \ \ \ \ \ \
j_i \equiv   \sql \pi_i \ , \ \ \ \ \ 
\pi_i=  {\tg_i\ov\tgg + \tg_1\tg_2\tg_3}  (\dot v_i  + 2 \J u_i) \ ,
\ee
where $u_3$ is such that  $\sumi \g_i u_i=0$
and we ignored terms linear in $\s$-derivatives 
of the fluctuations $v_i$ that integrate  to zero.
Thus  with the assumption that $\sum^3_{i=1} \tg_i\p_i$ does not
fluctuate  we see  that the  value of the total momentum 
$J= \sumi J_i$  is not, as required, 
 changed  by the fluctuations and thus the fluctuations of momenta 
 satisfy 
 $ \sumi j_i=0$.
 
To put the quadratic fluctuation Lagrangian into the canonical form it is useful
to do further  field redefinition to 4 fields $z_a,y_a$:
\be \la{yu}
v_1=  \sqrt{ {\D\ov 2(\tg_1 + \tg_3)  }} (   \sqrt{ {\tg_3 \ov \tg_1   }}
  z_1   -    \sqrt{ {\tg_2 \ov \tgg  }}   z_2 )  \ ,   \ \ \ \ \ \ \
  v_2 =  \sqrt{ {\D(\tg_1 + \tg_3) \ov 2 \tg_2 \tgg  }}    z_2 \ , \ee
\be   \la{op}
\ \ \
u_1= \sqrt{ {\tgg \ov 2  (  \tg_1 + \tg_3 )  } } (
\sqrt{ \tg_3 \ov \tg_1 }  y_1   -   \sqrt{  \tg_2 \ov \tgg
}     y_2 ) \ , \ \ \ \ \    u_2 =
  \sqrt{  \tg_1 + \tg_3 \ov 2\tg_2}   y_2 \ , \ \ \
\ee
where
\be\la{ty}
\D= \tgg + \tg_1 \tg_2 \tg_3 \ , \ \ \ \ \ \ \ \
\tgg= \tg_1 + \tg_2 + \tg_3 \ . \ee
Then the resulting  fluctuation action is
$
I=  {\sql}\int\, d\tau\int  {d\s\over
2\pi} \ L_2 $ where
(we assume summation over $a,b=1,2$)
\be \la{hj}
L_2 = \ha ( \dot y_a^2 - y_a'^2  +  \dot z_a^2 - z_a'^2 )
 - \ha B^2 y_a^2 +    A y_a \dot z_a
+    B \ep_{ab} y_b z'_a  \ , \ee
and
\be \la{er}
A=  2 \J  \sqrt{ \tgg \ov  \tgg + \tg_1 \tg_2 \tg_3  } \ , \ \ \ \ \ \
B=  2 \J \sqrt{ \tg_1 \tg_2 \tg_3\ov \tgg + \tg_1 \tg_2 \tg_3 }   \ , \ \ \ \ \ \ \
A^2 + B^2 = 4  \J^2 \ .
\ee
This quadratic  action can be
interpreted also as an action for a string in a  plane-wave
background in the l.c. gauge $x^+=\J \tau$ with $y_a,z_a$ representing
transverse coordinates. It  has constant coefficients and can be readily quantized
as discussed, e.g., in \ci{blau}.

For  $\g_i= \g$  \rf{hj}  reduces to  the fluctuation
Lagrangian found near $(J,J,J)$ geodesic in the LM case in \ci{mat,koch}.
In the case of $\g_i=0$ (when $A= 2 \J$) it  reduces to the BMN Lagrangian
in a rotated coordinate system
corresponding to the expansion near the  $(J,J,J)$ geodesic.

By writing down the corresponding equations of motion  and setting
$y_a, z_a \sim \sum_n   C_ n e^{i w_n  \tau + i n \s}$ one finds
that for $n\not=0$ the corresponding characteristic frequencies are
 the same as in the
BMN case, i.e. do not depend on $\g_i$ (for the  LM case  of equal $\g_i$
 this was  found in \ci{mat,koch}):
\be \la{fg}
w_n = \J   \pm \sqrt{ n^2 + \J^2 } = \J( 1 \pm \sqrt{ 1
+  { \tl n^2 }} ) \ .
\ee
Since we assumed that $t= \J \tau$, the corresponding fluctuation energies are
$E_n -\JJ = {|w_n| \ov \J}$. \foot{One may argue
that the conclusion that the spectrum
of fluctuations near this vacuum  $J_i \sim \g_i$ state
does not  depend on $\g_i$
 follows from the TsT construction  of this string theory in \ci{f}:
the difference in fluctuation spectra should involve $\nu_i$ and is
thus subleading in $1/\JJ$.
However, this does not  apply to the 0-modes, see below.}

%%%%%%%%%%%%%%%%%%%%%%%%%%%%%%%%%%%%%%%%%%%%%%%%%%%%%%%%
As was  shown  in \ci{mat}  in the case of $\g_i=\g$
the fermionic part of the  quadratic fluctuation Lagrangian
(in this case fermions are coupled to both the NS-NS and the R-R 3-forms
as well as  the R-R 5-form)
leads to the same spectrum as the bosonic Lagrangian, implying again that
there is a residual world-sheet supersymmetry (associated with supernumerary
\ci{pope}  target space supersymmetry). In particular,  the correction
to the ground state  energy cancels out.  This  should be  true also in the
present
unequal $\g_i$ case.
%NEW
\foot{String theory TsT relation suggests
 this  for integer $\nu_i$ in \rf{defo},
but cancellation between the bosonic and fermionic contributions
should not depend on whether $\nu_i$ is integer or not. This cancellation need
not persist at subleading orders when non-linear interactions of fluctuation
 modes are to be  included.}
The same bosonic spectrum will be found also on the
gauge theory side  from the analysis of the corresponding Bethe ansatz equations,
and  from the  coherent state action.

%%%%%%%%%%%%%%%%%%%%%%%%%%%%%%%%%%%%%%%%%%%%%%%%%%%%%%%%%%%%%%

\bigskip

The spectrum of the bosonic 0-modes (i.e. $\s$-independent fluctuations) is,
however,  nontrivial, as was already
 pointed out in the $\g_i=\g$ case in \ci{mat}.
The 0-modes  correspond to point-like strings,
i.e. represent fluctuations within the set of geodesics.
In the undeformed case all geodesics were BPS and thus had the same energy
the spectrum of 0-modes was degenerate; here this is no longer the
case (cf. \rf{par}).
The   Lagrangian for $\tau$-dependent 0-modes is
$L_2 = \ha ( \dot y_a^2  +  \dot z_a^2  )  - \ha B^2 y_a^2 +    A y_a
\dot z_a $
and this system  can be quantized by writing  down the corresponding
Schrodinger equation
and separating the oscillator dynamics (corresponding to $n=0$ case of
the above
$w_n$) from a free particle dynamics  as discussed,   e.g., in
\ci{blau}.
Same conclusion is reached
 also from the form of the
corresponding Hamiltonian (the momenta are $p_i=\sql \pi_i$):
$\H= \ha \pi^2_{y_a} + \ha (\pi_{z_a} - A y_a)^2 + \ha B^2 y^2_a$ 
(we omit the overall factor of string tension $\sql$).
Shifting  $y_a$ by $ - { A \ov A^2 + B^2} \pi_{z_a}$
 to isolate the oscillator dynamics in $y_a$-directions we end up with
\be \la{dina}
\H= \ha [\pi^2_{y_a}  + (A^2+ B^2) {\td y}^2_a ]
 + \ha { B^2\ov A^2 + B^2} \pi_{z_a}^2
= \ha ( \pi^2_{y_a}  + 4 \J^2  {\td y}^2_a     )
 + \ha {   \tg_1 \tg_2 \tg_3\ov  \tgg + \tg_1 \tg_2 \tg_3   }
\pi_{z_a}^2 \ . \ee
To express $\pi_{z_a}$ in terms of fluctuations of the angular
momenta  $j_i$ in \rf{jes} one should note that since  $\pi_{z_a}$ is given by a linear combination of $\dot
z_a $ and $ y_a$ (cf.\rf{hj}), redefining the latter
 by   $\pi_{z_a}$
changes also the relation between  $\pi_{z_a}$
and $\dot z_a $ (and thus commutation relations, etc.).
Equivalently, the same result for $H$ 
  is found in a more transparent 
way by  performing  the fluctuation analysis directly in 
the Hamiltonian for the $\s$-independent modes, i.e.  
by expanding both the coordinates and the momenta.
In terms of the fluctuations $\pi_i$  of 
the momenta of the angular coordinates in \rf{jes}, 
 the required phase-space 
  redefinition   
that separates the  oscillator dynamics 
 from the  free particle dynamics  is 
\begin{eqnarray}
u_1=
\frac{1}{2\J}\left[ 
\sqrt{\frac{\tg_3\tgg}{\tg_1(\tg_2+\tg_3)}}
\, y_1 -\sqrt{\frac{\tg_2}{\tg_2+\tg_3}} \, y_2 + 
{\frac{\tgg}{\tg_1} \pi_1}
\right] , 
\ \ \ \ 
u_2=\frac{1}{2\J}\left[\sqrt{\frac{\tg_1+\tg_3}{\tg_1}}
\, y_2 + \frac{\tgg}{\tg_1} \pi_2 \right]  
\end{eqnarray}
This leads to  the Hamiltonian 
\begin{eqnarray}
\H=\ha\left(\pi^2_{y_a}+4{\cal J}^2 y_a^2\right)+
\frac{1}{2\J}\bigg[  \tg_2  (  \tg_1 + \tg_3)
\pi_1^2  + 2 \tg_1 \tg_2 \pi_1 \pi_2
+   \tg_1  (  \tg_2 + \tg_3) \pi_2^2 \bigg] \ . 
\end{eqnarray}
Thus the 0-mode contribution to the fluctuation energy spectrum 
 expressed in terms of the angular momenta
  of the fluctuation modes in \rf{jes}
\be \la{moo}
j_i= J_i- {\g_i \ov \gga} \JJ\ ,\  \ \ \ \ \ \ \ \ \sumi j_i =0 \ ,
   \ee
   takes the form ($E=\sql \H$)
\be \la{spe}
E_{0-mode}  
= { \l \ov 2 \JJ} \bigg[
  \g_2  (  \g_1 + \g_3)
j_1^2  + 2 \g_1 \g_2 j_1 j_2
+   \g_1  (  \g_2 + \g_3) j_2^2 \bigg] \ . \ee
In the case when $\g_i = \g$  eq.\rf{spe}  becomes simply
\be \la{ti}
E_{0-mode} =
{ \l \g^2 \ov  \JJ} \big[ (J_1 - { 1 \ov 3}  \JJ)^2 + (J_2 - { 1 \ov
3}  \JJ)^2
+  (J_1 - { 1 \ov 3}  \JJ)  (J_2 - { 1 \ov 3}  \JJ)  \big] \ , \ee
reproducing
the expression in  \ci{mat}
%(up to a sign of the last term)
\be\la{ooo}
E_{0-mode} =
{ \l \g^2 \ov  3 \JJ} \big[ (J_1 - J_2 )^2 + (J_1 - J_3 )^2
    - (J_1 - J_2) (J_1-J_3)   \big] \ . \ee
 Let us stress 
%make a general remark  
that the correct   quantization of
the zero mode sector
should not be based on the  expansion to quadratic  order in fluctuations
but should start directly with the (supersymmetric version of) the
Neumann model \rf{par},  i.e. from the corresponding 0-mode truncation of
the superstring action. In the undeformed \adss case this amounts to
quantizing the
corresponding superparticle action leading to the spectrum
 of the BPS (supergravity) modes.
The first attempt in this direction would be
 to keep only the bosonic
fields (and thus ignore the ``mixing'' with the $AdS_5$ directions via fermions)
and try to use the known information about quantum Neumann model
(see, e.g., \ci{quant}). Such  0-mode sector quantization of this $\g_i$
deformed string theory would be equivalent (for integer $\nu_i$)
 to  a ``minisuperspace'' quantization of
the original \adss  string theory in the sector of rotating and
pulsating circular strings.

Assuming  the large $\JJ$ limit to suppress  quantum corrections,
the   expression \rf{spe} can be found directly from the
particle Hamiltonian \rf{par},\rf{jo} by considering   the
 semiclassical configurations
 with  constant $\r_i$ (to  minimize energy  for given $J_i$).
 In the limit of large $\J_i \sim \J$ with
 fixed $\nu_i=\epsilon_{ijk} \tg \g_j \J_k$
 it is sufficient to  evaluate the   energy on the
 semiclassical configuration that extremises the dominant
 $S^5$ part of  potential in  \rf{par}:
 \be \la{rhoo}
 \rho^2_i = {J_i \ov \JJ} \ .
\ee
 For zero $\g_i$, i.e. for the semiclassical
particle states  represented by
 geodesics on $S^5$  these are BPS states of undeformed theory.
 For $\g_i\not=0$ if  $J_i$ happen to be equal
  to ${\g_i \ov \gga} \JJ$
 these  are   the vacuum states of deformed theory  
 having again   $E= \JJ$.
  The  energy of the   configuration \rf{rhoo} is
  $E= \sql \sqrt V $,   where $V = \sumi V_i$ is the potential
 in \rf{par} evaluated on $\r_i$ in  \rf{rhoo}, i.e.  
 \be \la{genr}
 E= \sqrt{ \JJ^2 + \l \bigg[
 \g_2  \g_3  j_1^2  +   \g_1  \g_3 j_2^2 +   \g_1 \g_2 j^2_3
 - { 1 \ov \JJ} (\g_1 +\g_2 + \g_3)^2 j_1 j_2 j_3 \bigg] } \ , \ \ \ \ \
 j_i \equiv J_i -  {\g_i \ov \gga} \JJ\ .     \ee
 Here in general  $j_i$ or $J_i$
  are of the same order as $\JJ=J_1 + J_2 + J_3 $.
 Expanding \rf{genr}   in  large $\JJ$ 
 for fixed $  \g j$ and $\tl = {\l \ov \JJ^2}$
  we get, as for semiclassical string states,
  to leading order in $\tl= { 1 \ov \J^2}$
  %$E=\JJ $ plus the  correction in \rf{spe}.
\be \la{spi}
E= \JJ + { \l \ov 2 \JJ} \bigg[ (\g_2  \g_3  j_1^2  +
  \g_1  \g_3 j_2^2 +   \g_1 \g_2 j^2_3) -
  { 1 \ov \JJ} (\g_1 +\g_2 + \g_3)^2 j_1 j_2 j_3    \bigg]  +
   O({\l^2 \g^4 j^4\ov \JJ^3} )
   \ .\ee
The  $j^2$ term here is the same as in \rf{spe}.
We shall reproduce this expression  from the
fast string limit  in section 3 or from the
Landau-Lifshitz model in section 5.
We shall also  obtain the same
spectrum of   0-mode fluctuations \rf{spi}
 on the gauge theory side directly from the Bethe ansatz
 in Appendix B.

Let us note that in  the limit which we consider
($\JJ \to \infty  $ with  $\tl ={ \l\ov \JJ^2} $
 and $ \JJ \g_i$ fixed)
% it may seem natural to demand that the contribution
the  $j^2$  contribution in \rf{spe}
 is finite
 %(like  other contributions in the BMN spectrum)
provided $ j_i \sim  \JJ^{1/2}$
  (or, for $\g_i=\g$,  if $J_i-J_j \sim  \JJ^{1/2}$ \ci{mat}).
However, as already stressed  above, 
 the true condition of validity of \rf{spe}
 follows from the exact treatment of the 0-mode fluctuations
 and is only
that $j_i \sim  \JJ^\mu , \  \mu \leq  1. $
%to justify the expansion
%of \rf{spi}  in powers of $j_i$ at large $\JJ$
The same condition will
appear  on the spin chain side in Appendix B.
As for the  quantum  corrections,
they are expected to modify \rf{spi} by  terms
 with similar  dependence on
$j_i$  but suppressed by extra  powers of $1/\JJ$.
 Same   structures should   appear
on the spin chain side as  corrections to leading
 thermodynamic limit
approximation.

\bigskip

%NEW

\renewcommand{\theequation}{3.\arabic{equation}}
 \setcounter{equation}{0}

%%%%%%%%%%%%%%%%%%%%%%%%%%%%%%%%%%%%%%%%%%
\section{Fast motion limit: Landau-Lifshitz action from the string action}
%%%%%%%%%%%%%%%%%%%%%%%%%%%%%%%%%%%%%%%%%%%%%%%%%%

Let us start with recalling the  derivation of the
reduced effective action that governs the dynamics of ``slow'' string degrees of freedom
in the 3-spin ($su(3)$ invariant)  sector of undeformed theory  following \ci{kt}.
We shall  parametrize $S^5$ by 3 complex coordinates $X_i$ such that
\be \la{parr}
X_i = \r_i e^{i\p_i} \equiv  U_i e^{i \psi} \ , \ \ \ \ \ \ \ \ \ \  \sumi \r^2_i=1  \ , \ee
where $\r_i$ and $\p_i$ are real. We have isolated the common
phase $\psi$ that  will be a  collective coordinate representing fast string motion
in the three planes. There is an obvious $U(1)$ gauge invariance
$U_i \to e^{i \zeta  } U_i, \ \psi \to \psi - \zeta$.
The $S^5$ metric  has then the form of the Hopf fibration of $S^1$ over $\IC\IP^2$:
\be
ds^2 = dX_i dX^*_i =\sumi ( d\r_i^2 + \r^2_i d\p_i^2) =
  (d \psi + C)^2 +  dU_i^* dU_i - C^2\ , \ee
where $C= - i U^*_i d U_i$.  The $\IC\IP^2$ metric is $dU_i^* dU_i  + ( U^*_i d U_i)^2
  = | D U_i|^2 $
where $DU_i= dU_i - i C  U_i$. We can then start with  the general form of the
 bosonic part of the string action  and  apply
 the 2-d duality (T-duality) in the $\psi $
direction. The result is (including time direction of $AdS_5$ and assuming summation over i)
\be
L=  \ep^{pq} C_p \d_q \tilde \psi
- \ha \sqrt{- g} g^{pq}  ( - \d_p t \d_q t  + \d_p \ta \d_q \ta  + D_p U_i^* D_q U_i)
\ , \ee
where the first term represents  the 2-form coupling induced  by
 off-diagonal form of the metric. The next step is solve for the 2-d metric,
replacing the second term
in $L$ by its Nambu counterpart, $\sqrt{ - \det h}$, $\ \
h_{pq} = - \d_p t \d_q t  + \d_p \ta \d_q \ta  + D_p U_i^* D_q U_i$. The final
 step is  to fix a static gauge: $t= \tau, \ \ta= \J \s$, where the letter
condition corresponds to fixing the angular momentum  associated with
the fast variable $\psi$, i.e. the total momentum $\JJ=J_1 + J_2 + J_3$,
to be homogeneously distributed  along $\s$ (as this is the property
of the spin chain description of the corresponding states). Finally,
expanding in large $\J$ and assuming  that time derivatives of
the slow ``transverse'' variables  $U_i$ are small we end  with
the following $\IC\IP^2$ analog of the Landau-Lifshitz
 action ($\tl = {\l \ov \JJ^2}$):
\be \la{lal}
I= \JJ \int dt \int^{2\pi}_0 { d \s \ov 2 \pi}  \  [ 
 \L + O(\tl^2) ] \ , \ \ \ \ \ \ \ \ \ \ 
\L= - i  U^*_i \d_t U_i  -  \ha \tl  | D_\s U_i|^2  
  \ .   \ee
Since this action has $U(1)$ gauge invariance (in addition to the
global $SU(3)$ invariance), we may parametrize
$U_i$ in the same  way as in \rf{par}, i.e.
$U_i = \r_i e^{i\p_i} $ ($\sumi \p_i$ can be assumed to be gauge fixed to
 zero but will in any case
 decouple) getting explicitly
\be \la{eex}
\L=  \sumi \r^2_i \dot \p_i  -  \ha \tl  \bigg[
 \sumi \r'^2_i  +
 \sum^3_{i<j=1} ( \p_i'-\p'_j)^2 \r^2_i \r^2_j  \bigg] \ . \ee
Note that since $\sumi \r^2_i=1$ the  WZ term depends
(modulo a total derivative) only on $\p_i-\p_j$, i.e.  $\sumi \p_i$ indeed
decouples.
Other forms of this action were given in \ci{hlo,st}.
This $\IC\IP^2$ action is integrable, i.e. the corresponding equations of motion
admit a Lax pair representation.
\foot{To find the corresponding  Lax representation
it is useful  to use the  matrix form of the  LL model
\ci{st}  in which the LL equation takes the form (we rescale time to absorb $\tl$):
             $\d_t  N = - {i\ov 6} [ N, \d^2_\s  N]$. Here
$N_{ij} = 3 U^*_i U_j -  \delta_{ij}$
satisfies $
\Tr N=0,\  N^\dagger =\   N^2 = N +2$.
This equation  can be written as
$     \d_t N = \d_\s K$ ,
 where $K\equiv  - {i\ov 6}  [ N, \d_\s N]$.
We observe that $N$ is ``covariantly constant''
        $\d_\s N = { 2i \ov 3 } [N,K]$
     and define the Lax connection $(A_\s,A_t)$
     as
$A_\s=  is N,  \   A_t =  is K + {3i\ov 2} s^2 N$
($s$ is a spectral parameter).  It then
satisfies  (as a consequence of the above two equations on $N$ and $K$)
  $ \d_t A_\s - \d_\s A_t - [A_t,A_\s]=0$.}

\bigskip

\subsection{Deformed case}

%%%%%%%%%%%%%%%%%%%%%%%%%%%%%%%%%%%%%%%%%%%%%%%%%
Let us now  find a generalization of this  $\IC\IP^2$  action to the case of
non-zero deformation parameters $\g_i$. We may choose
$\psi= \sum_i \p_i$\  (e.g., set
$\p_1 = \psi - \vp_2, \ \p_2 = \psi + \vp_1 +\vp_2, \ \p_3 = \psi - \vp_1$)
 and start  with the metric and $B_2$ background in \rf{me},\rf{bfield0}.
After doing T-duality and the same  gauge fixing as above we finish with
 the following  generalization of \rf{eex}
\be \la{eexx}
\L=  \sumi \r^2_i \dot \p_i  -  \ha \tl   \bigg[
 \sumi \r'^2_i  +
 \sum^3_{i<j=1} ( \p_i'-\p'_j -  \ep_{ijk} \bg_k )^2 \r^2_i \r^2_j
-  \bgg^2  \r^2_1 \r^2_2 \r^2_3    \bigg]  \ ,   \ee
where
\be \la{ba}
\bg_i  \equiv \tg_i \J = \g_i \JJ \ , \ \ \ \ \ \ \ \ \ \ \ \ \
\bgg\equiv \sumi \bg_i   \ . \ee
Note the the  time-derivative   (WZ)
 term does not get deformed.

Since the 3-parameter deformation of the full string model is integrable \ci{f},
this action should represent an integrable deformation
of the $\IC\IP^2$ LL model.\foot{Indeed, this action describes (a large $\J$)
 approximation to solutions  of the original string action.
It would be interesting to find explicitly
the corresponding Lax pair.}

The 2-spin case action  is recovered by setting $\r_3=0$;
then  $\r_1^2 + \r_2^2=1$ and  the action depends only on
 $\g_1$ and reduces to the anisotropic version of the  $\IC\IP^2$ LL model
found  in \ci{frt}.

We observe  that the case of \rf{eexx} with  $\tgg=0$ is special:
then the dependence on $\bg_i$  can be formally absorbed
 into
a formal redefinition of $\p_i$  (as was the case in the 2-spin sector in \ci{frt}),
 e.g., $\p_1 \to \p_1 + \bg_3 \s, \
\p_2 \to \p_2 , \  \p_3 \to \p_3 - \bg_1  \s$;
in terms of the shifted angles the action then becomes the same as \rf{eex}.

Another special case is  the symmetric one
 $\g_i=\g$ when we get explicitly
\be \la{rexx}
\L=  \sumi \r^2_i \dot \p_i  -  \ha \tl \H \ , \ \ \ \ee
\be \la{exx}
 \H=
 \sumi \r'^2_i  +
 ( \p_1'-\p'_2 - \bg  )^2 \r^2_1\r^2_2 +
( \p_2'-\p'_3 - \bg  )^2 \r^2_2\r^2_3 +
( \p_3'-\p'_1 -  \bg  )^2 \r^2_3\r^2_1
-  9 \bg^2  \r^2_1 \r^2_2 \r^2_3     \ .  \ee
Another   generalization of \rf{exx}  (``orthogonal'' to the one in  \rf{eexx})
  can be found by starting  with more a general  supergravity background in  \ci{LM}
   dual to $\N=4$ SYM deformation with complex parameter
$\b=\g - i\bfsigma $. That background depends on both
$\tg = \sql \g$ and $\tk = \sql \bfsigma$  and was
 obtained in \ci{LM} using S-duality
transformations.\foot{For comments on
the corresponding string theory  and  AdS/CFT duality in this case
  see also \ci{frt}.}
In this case  the application of the above  procedure  leads to \rf{exx}
with $\H$ replaced by \foot{In this case $B_2$ has a term proportional to $w_1$
and one is to use \rf{w1}.}
$$
\H= ( \r_1 \r_2' - \r_2 \r_1' + \bk \r_1\r_2)^2
+ ( \r_3 \r_1' - \r_1 \r_3' + \bk \r_3\r_1)^2
+( \r_2 \r_3' - \r_3 \r_2' + \bk \r_2\r_3)^2 $$ \be
 \la{ey}
+
 ( \p_1'-\p'_2 - \bg  )^2 \r^2_1\r^2_2 +
( \p_2'-\p'_3 - \bg  )^2 \r^2_2\r^2_3 +
( \p_3'-\p'_1 -  \bg  )^2 \r^2_3\r^2_1
-  9 (\bg^2 + \bk^2)  \r^2_1 \r^2_2 \r^2_3     \ ,
 \ee
where $\bk \equiv  \tk\J  = \bfsigma  \JJ $.
This deformation of the $\IC\IP^2$ action \rf{eex} is unlikely to be
 integrable.
The action \rf{eexx} admits a similar generalization to the case of
the 3 different $\bk_i$ parameters.

%%%%%%%%%%%%%%%%%%%%%%%%%%%%%%%%%%%%
\subsection{Special solutions}
%%%%%%%%%%%%%%%%%%%%%%%%%%%%%%%%%%%%%

Let us now study some solutions  of the action \rf{eexx}.
The solutions are characterized by 3 conserved angular momenta
\be \la{ja}
J_i = \JJ \int^{2\pi}_0 { d \s \ov 2 \pi}\ \r^2_i \ , \ \ \ \ \ \ \
 \JJ= J_1 + J_2 + J_3 \ .  \ee
The  main  difference between  \rf{eexx}
 and its  undeformed case \rf{eex}
is the presence of the (non-negative) potential term:
\be \la{poo}
V= \bg_3 ^2 \r^2_1\r^2_2 +
   \bg_1^2 \r^2_2\r^2_3 + \bg^2_2 \r^2_3\r^2_1
-   (\bg_1 + \bg_2 + \bg_3)^2  \r^2_1 \r^2_2 \r^2_3
  \ ,  \ee
where $\r_i$ are subject to  $\sumi\r^2_i =1 $.
While in the case of $\g_i=0$ the action \rf{eex}  had $\s$-independent
solutions
$\p_i=\const, \ \r_i=\const$ (with $\r_i$ being arbitrary apart
from $\sumi \r^2_i=1$) which represented  BPS geodesics  with $E=\JJ$
now the potential selects  only few of them that will
minimize $V$ and thus  $E=\JJ (1 + \ha \tl  V )$,
i.e. will  have $V=0$. These absolute minima of $V$   correspond precisely
to the BPS geodesics discussed above in section 2,  namely,
\be
(i)\  \r_1=1, \ \r_2=\r_3=0 ; \ \ \  \r_2=1, \ \r_1=\r_3=0 ;  \ \ \
\r_3=1, \ \r_2=\r_1=0,\ee \be \la{ghe}
(ii)\ \r_i = \sqrt{ \bg_i \ov \bgg}, \ \ \ \ \ \ {\rm i.e.}\ \ \
 \ J_i = {\g_i\ov \gga} \JJ \ . \ee
Other geodesics described (in this large $\J$ approximation)
by solutions with  constant values of $\r_i$
(note that for $\s$-independent  solutions
 $\p_i-\p_j$ play in \rf{eexx} the role of the Lagrange multipliers
 imposing the condition of constancy of $\r_i$)
will have non-zero value of the energy.
Explicitly, we find from \rf{poo} that  for generic point-like solutions
\be \la{eewi}
E= \JJ +  \ha   \tl  \bigg[ { 1\ov \JJ}
( \bg_3 ^2{ J_1 J_2 }  +
   \bg_1^2 J_2 J_3 + \bg^2_2 J_3  J_1 )
-    { 1\ov \JJ^2 } (\bg_1 + \bg_2 + \bg_3)^2  J_1 J_2 J_3 \bigg]
\ . \ee
For fixed $\bg_i$ such  dependence of energy on spins $J_i$ is characteristic of
macroscopic string solutions \ci{ft2}.
%We should note that here we are expanding at large $\J$ so $J_i-J_j$
%are small and it is natural to express the energy in terms of these
%differences.
 An alternative representation making it clear that the energy vanishes
for  $J_i  \sim \g_i$ states is (cf. \rf{spi}) 
\be \la{eew}
E= \JJ +  \ha   \tl  \bigg[ { 1\ov \JJ}
[  \bg_2 (\bg_1 + \bg_3) j_1 ^2 +  2 \bg_1 \bg_2 j_1 j_2
+  \bg_1 (\bg_2 + \bg_3) j_2 ^2]
-    { 1\ov \JJ^2 } (\bg_1 + \bg_2 + \bg_3)^2  j_1 j_2 j_3 \bigg]
\ ,  \ee
where as in \rf{moo}
$j_i = J_i - { \bg_i \ov \bg_1 + \bg_2 + \bg_3} \JJ$.
In the LM  case  of $\bg_i=\bg$ this can be written also as
\be \la{eeuw}
E= \JJ +     \tl \bg^2  \bigg[ { 1\ov \JJ}
( j_1 ^2 +  j_1 j_2 +  j_2    ^2) -    { 9\ov 2 \JJ^2 }   j_1 j_2 j_3 \bigg]
\ , \ \ \ \  \ \ \   j_i = J_i - \thr  \JJ \ ,  \ee
where the $j^2$ term is recognized  to be equivalent to the 0-mode contribution
in \rf{ti} or \rf{ooo}. 
Thus the zero-mode contributions are easily captured by the LL model.

If  $\tl,\ \bg$ and $j^2/\JJ$ are fixed
as one may   assume in the discussion of 
the quadratic  zero-mode fluctuation contribution to the energy
spectrum near the $(J,J,J)$ geodesic, then $j^3/\JJ^2$ term
 is subleading. As already mentioned in section 2.3.2, 
 the assumption that $j^2/\JJ$ is fixed is not needed in general, 
 and in Appendix B we shall reproduce the whole 
 expression \rf{eew} including the $j^3$ term 
  from the Bethe ansatz on the gauge theory 
 side.

It is also straightforward to study the non-zero part
of the fluctuation spectrum near  $(\JJ,0,0)$ or
$J_i\sim \g_i$ geodesic  and to  show that it is agreement with the leading order $\tl$ term
in the corresponding  part of   spectrum found in sections
 2.3.1 and  2.3.2.\foot{One can also study  fluctuations near more general geodesics,
and here the spectrum will be similar to the one found in  \ci{ft3,art}
near circular rotating strings
in undeformed theory.}

Let us now comment on  extended string  solutions of the LL action.
One observes that in general the circular string  ansatz
\be \la{cii}
\phi_i = m_i \s \ ,  \ \ \ \ \ \  \   \ \ \  \r_i=\const \ ,  \ee
gives a solution  of  LL action \rf{exx} with $\g_i=\g$.
Similar circular solutions exist in the full string
equations  and are  analogs of the rigid circular strings in
 undeformed \adss geometry \ci{ft2,art}.\foot{Some 
 string solutions in LM geometry were discussed in \ci{bob}.}
 Interestingly, as it is obvious from the
comparison of \rf{exx} and \rf{poo},
this case is formally equivalent to the case
of {\it point-like}  solutions in the  $\g_i$-deformed  theory with
\be \la{gag}
\g_1 = \g + { m_3-m_2  \ov \JJ}\ , \ \ \ \
\g_2 = \g + { m_1-m_3  \ov \JJ}\ , \ \ \ \
\g_3 = \g + { m_2-m_1  \ov \JJ}\ ,  \ \ \ \ \  \sumi \g_i = 3 \g \ .  \ee
In this case we know that there are vacuum $(J_1,J_2,J_3)$ states
 provided
$J_i= {\g_i\ov  \g_1 + \g_2 + \g_3 }   \JJ$.  Here that leads to
 the conclusion
that we can have ground-state solutions  provided
$\g$ takes special  rational  values.
These are, in fact, the string  BPS states found  in the
supersymmetric deformed theory in \ci{LM}. They  are  TsT images of
particular point-like BPS   states in the undeformed \adss theory
(there perturbative large $N$
 BPS states are represented only by  point-like strings).

Similar remark applies if we start with  the  generic $\g_i$ case of LL action
\rf{eexx}: specifying to the  sector of  circular  strings with constant
 ``radii'' $\r_i$
is equivalent to studying  point-like states in the theory with
shifted $\g_i$ parameters: $\g_1 \to \hat \g_1= \g_1 + { m_3-m_2
 \ov \JJ}, $ etc.
One then finds additional ground states for
special values of $J_i$ and $\g_i$  such that
$\ep_{ijk} J_j \hat  \g_k$ are zero.

% i.e. for such
%(integer) $J_i$ and (rational) $\g_i$ that their vector product is integer,
%\be \la{rat}
%\ep_{ijk} J_j  \g_k =   m_i \ . \ee

\renewcommand{\theequation}{4.\arabic{equation}}
 \setcounter{equation}{0}

%%%%%%%%%%%%%%%%%%%%%%%%%%%%%%%%%%%%%%%%%%%%
%%%%%%%%%%%%%%%%%%%%%%%%%%%%%%%%%%%%%%%%%%%%

%%%%%%%%%%%%%%%%%%%%%%%%%%%%%%%%%%%%%%%%%%%%%%%%%%%%%%%%%%%%%%%%%%%%%%
\section{Dilatation operator  of deformed  gauge theory}
%%%%%%%%%%%%%%%%%%%%%%%%%%%%%%%%%%%%%%%%%%%%%%%%

%%%%%%%%%%%%%%%%%%%%%%%%%%%%%%%%%%
\subsection{The spin chain Hamiltonian for the holomorphic 3-scalar sector}
%%%%%%%%%%%%%%%%%%%%%%%%%%%%%%%%%%%%%%%%%%%%%%%%%

The one-loop planar  dilatation operator of the  $\N=4$ SYM theory
in the  holomorphic 3-scalar  sector  (i.e. the anomalous dimension
matrix for  the operators $\Tr (\P^{J_1}_1  \P^{J_2}_2 \P_3^{J_3} +...)$
 built out of chiral scalars $\P_i$, $i=1,2,3$)
can be written as an  $su(3)$ invariant nearest-neighbor ferromagnetic
spin chain  Hamiltonian  \ci{mz}:
\be \la{hah}
H=  \sum_{k=1}^L H_{k,k+1} \ ,\ \ \ \ \
 \ \ \ H_{k,k+1}= \frac{\lambda}{8\pi^2}\H_{k,k+1}\ ,  \ee
\be \la{haoo}
 \H^{(0)}_{k,k+1} \equiv  \id_{k,k+1} - \cP_{k,k+1} \ . \ee
Here $H$ acts on products of 3-vectors at each site
of the spin length $L$  which is equal to the total momentum
(in discussion of spin chains and Bethe ansatz we shall use the notation $L$
instead of $\JJ$)
\be \la{lll}
L = \JJ \equiv  J_1+J_2+J_3 \ . \ee
 $\id$ is
an  identity  and $\cP$ is the permutation operator.
In terms of
the generators $(e_n^m)_i^j \equiv  \delta_i^m\delta_n^j$
 of the algebra $gl(3)$ we have
\begin{eqnarray} \la{yy}
\id_{k,k+1} = \id_k\otimes\id_{k+1}=  \sumn e^m_m(k) e^n_n (k+1) \ , \
\ \ \ \ \ \
\cP_{k,k+1} = \sumn e^m_n(k) e^n_m (k+1)\  .
\end{eqnarray}
The generalization of \rf{hah} to  the case of the $\b$-deformed
$\N=4$ SYM theory  was found in \ci{RR,BECH}.
It has a formal  generalization to the case of 3 complex
deformation parameters
$q_i= e^{i \pi \b_i}$ (here $e\otimes e \equiv e(k) e(k+1)$):
\begin{eqnarray}
\H_{k,k+1} = &&
 |q_1|^{-2}e_2^2\otimes e_3^3+|q_1|^{2}e_3^3\otimes e_2^2
-\frac{q_1}{{\bar q}_1}e_2^3\otimes e_3^2
- \frac{{\bar q}_1}{q_1}e_3^2\otimes e_2^3\cr
&+& |q_2|^{-2}e_3^3\otimes e_1^1+|q_2|^{2}e_1^1\otimes e_3^3
-\frac{q_2}{{\bar q}_2}e_3^1\otimes e_1^3
- \frac{{\bar q}_2}{q_2}e_1^3\otimes e_3^1\cr
&+& |q_3|^{-2}e_1^1\otimes e_2^2+|q_3|^{2}e_2^2\otimes e_1^1
-\frac{q_3}{{\bar q}_3}e_1^2\otimes e_2^1
- \frac{{\bar q}_3}{q_3}e_2^1\otimes e_1^2~~ . \la{hi}
\end{eqnarray}
This expression   appeared in
\cite{RR} as a step in the construction of the Hamiltonian for the
supersymmetric deformation in the 3-spin sector which
 corresponds to
equal parameters $q_i=q=e^{i\pi \b}$.\foot{In the general complex
$\b$ case one needs also a rescaling of the coefficient $\l$
in front of $\H_{k,k+1}$ as discussed in \ci{frt}.}
It was noticed in \ci{RR} that the complex $\b$   deformation is not
contained in the class of integrable deformations of
$su(3)$-invariant Heisenberg chain described by a twisted $R$-matrix.
It was later argued \cite{BECH} that the spin chain describing
the complex $\b$  case is not integrable.

In the case of real $\b_i\equiv \g_i$ which we will be interested in here
\rf{hi}   becomes
\be \la{hp}
\H_{k,k+1} = \id_{k,k+1} - \td \cP_{k,k+1} \ , \ \ \ \ \ \ \ \ \ \ \
\td \cP_{k,k+1} = \sumn e^{-2i \pi \a_{nm}}  e^m_n(k) e^n_m (k+1) \ ,\ee
\be \la{alp}
\a_{mn} \equiv -\ep_{mni} \g_i \ . \ee
This  gives the 1-loop dilatation operator of the non-supersymmetric
deformation of $\N=4$  SYM theory  \ci{f}
which should be dual  to  string theory defined by \rf{me}--\rf{w1}.
This gauge theory has  the following  scalar potential \ci{f}
\be \la{poi}
V ={ \rm Tr} \sum^3_{n> m=1}
%  | e^{-i\pi \a_{mn}} \P_n \P_m -
% e^{i\pi \a_{mn} }  \P_m \P_n  |^2    +
 | e^{-i\pi \a_{mn}} \P_m \P_n -
 e^{i\pi \a_{mn} }  \P_n \P_m  |^2    +
 {\rm Tr} \sum^3_{m=1}
[ \P_m ,  \bar \P_m]^2 \
\ee
and similarly deformed  Yukawa couplings
to ensure
the marginality of the deformation as well as the
cancellation of the self-energy corrections
to the anomalous dimension matrix.

The terms displayed in (\ref{hp}) are determined by the interactions
in the first sum in \rf{poi}; for (\ref{hp}) to be indeed the dilatation
operator it is necessary that, for general $\gamma_i$,
the contribution of self-energy graphs, vector exchange graphs and the
graphs containing $[\Phi_m, \bar \Phi_m]^2$ interaction vertices
continue to cancel out, just like in the supersymmetric theory case.
 This cancellation is relatively easy to understand based on the
similarity between the $\beta$ deformation and noncommutative
theories \ci{BELE,LM}: here  we have a noncommutative structure
related to  the $U(1)$ symmetries inherited from the R-symmetry of the
undeformed theory. In noncommutative theories, planar graphs in
the deformed theory are equal \ci{filk} to those in the undeformed
theory except that the  external fields are multiplied with a
*-product.
%%%%
%% additions below %AT
%%%%
\foot{The noncommutativity due to the $\g_i$-deformation is certainly
different from the one discussed in \ci{filk}, so one may be inclined
to question the applicability  of the results of \ci{filk}
 to our case. Abstractly, the results of \ci{filk} are based on the fact
that the fields  carry certain additive charges and that the
corresponding symmetry generators obey the chain rule. These
are the  properties of the $U(1)$ symmetries inherited from the R-symmetry
of the $\N=4$ SYM theory in the present case  as well
 as the momentum generators in   the case of
\ci{filk}. The difference between the two cases 
 is that, while all the fields
carry momentum, in our 
 case of the $\g_i$-deformed theory  some fields 
 have trivial charges so are
not affected by the *-product. We will return to 
this point at the end
of this section.
}
%%%%
%% additions above
%%%%
The cancellation mentioned above occurs as follows.
The  vector exchange graphs are independent of the deformation because
the  vector-scalar-scalar coupling is  independent of  $\g_i$. Similarly,
the vertices analogous to those coming from the ``D-term''
%the Kahler potential
 $[\Phi_m, \bar \Phi_m]^2$   are undeformed because the
charge vectors of $\Phi_m $ and $\bar \Phi_m$
are proportional.
The deformation  of the Yukawa couplings  is done again with the
*-product which now contains the fermion $U(1)$-charges (equal to
their R-charges in the undeformed theory).
The contribution of fermions to self-energy thus may  have  a
nontrivial phase, but,  based on the noncommutative structure, the
planar self-energies would be the same as in  the
$\N=4$ theory  except for  a
*-product between the external  fields. Due to the $U(1)$-charge
conservation,  this phase is 1, i.e. there is
no nontrivial contribution.
 As a result,
 there is the same cancellation as in the $\N=4 $ theory
between the vector exchange, self-energy and the contribution of the
  $[\Phi_m, \bar \Phi_m]^2$  vertices, as it should be
for \rf{hp}  to correspond to the one-loop
dilatation operator of the gauge theory deformation suggested in \ci{f}.

\bigskip

Let us note that for the  non-holomorphic sectors of the
$\b$ deformed theory it is complicated to construct the
dilatation operator  by a  direct computation, even
in the case of real deformation. Using different techniques, it was
shown in \cite{BR} that, for any sector, the Hamiltonian of the spin chain
in the deformed theory $H$ is related to the Hamiltonian in the
undeformed theory $H^{(0)}$ by
\begin{eqnarray}
H_{k,k+1} &=&  \cU_{k,k+1} H^{(0)}_{k,k+1}  \cU^{-1}_{k,k+1}  \ , \ \ \ \ \ \  \
\cU_{k,k+1}=
e^{i\pi\sum_{m,n=1}^3 \alpha_{mn} \hh_m(k)\hh_n(k+1) } \ ,
\label{Hstar}
\end{eqnarray}
where $\hh_n(k)$ are the Cartan generators of the symmetry group acting
at site $k$.
In  the case of our  present interest, i.e.
 the holomorphic 3-scalar sector
\be \la{vv}
\H_{k,k+1} =
\cU_{k,k+1}
\left(\id_{k,k+1}-{\cal P}_{k,k+1}\right)
\cU^{-1}_{k,k+1} \ , \ee \be \la{uu}
\cU_{k,k+1}=  e^{i\pi\sum_{m,n=1}^3 \alpha_{mn} e_m^m(k)e_n^n(k+1)}
= \sum_{m,n=1}^3 e^{ i\pi\alpha_{mn}  }   e_m^m(k)e_n^n(k+1) \ ,
\label{HstarSU3}
\ee
where we used \rf{yy} and  that
$e^n_n (k) e^m_m (k) = \delta^m_n  e^m_m (k)$.

%%%%%%%%%%%%%%%%%%%%%%%%%%%%%%%%%%%%%%%%%%%%%%%%%%
\subsection{The Bethe ansatz}
%%%%%%%%%%%%%%%%%%%%%%%%%%%%%%%%%%%%%

As usual, the diagonalization of a spin chain Hamiltonian with more than two
states per site is done through the nested Bethe ansatz algorithm.
{}From the details described in \cite{RR} it is straightforward though
tedious to derive the Bethe equations for the 3-spin sector; one  can
also specialize the results of \cite{BR} to this case. The resulting  Bethe equations
are
\begin{eqnarray}
&&\!\!\!\!
e^{-2i\pi L\alpha_{21}}\left[\frac{u_{1,k}+\ialf}
{u_{1,k}-\ialf} \right]^{L}
=
\prod_{\stackrel{i=1}{i\ne k}}^{J_2+J_3}
\frac{u_{1,k}-u_{1,i}+i}
{u_{1,k}-u_{1,i}-i}
\left[
\prod_{{j=1}}^{J_3}e^{-
2i\pi (\alpha_{32}+\alpha_{21}+\alpha_{13})}
\frac{u_{1,k}-u_{2,j}-\ialf}
{u_{1,k}-u_{2,j}+\ialf}
\right]
\label{mainBetheSU3}\\
&&\!\!\!\!
e^{2i\pi L(\alpha_{21}+\alpha_{13})}
=
\prod_{\stackrel{j=1}{j\ne l}}^{J_3}
\frac{u_{2,l}-u_{2,j}+i}
{u_{2,l}-u_{2,j}-i}
\left[\prod_{i=1}^{J_2+J_3}
e^{2i\pi(\alpha_{32}+\alpha_{21}+\alpha_{13})}
\frac{u_{1,i}-u_{2,l}+\ialf}
{u_{1,i}-u_{2,l}-\ialf}\right]\ .
\label{auxBetheSU3}
\end{eqnarray}
Here $L=J_1 + J_2 + J_3$ and we should add also
 the condition that the eigenvectors are related to
single-trace operators (the cyclicity condition):
\begin{eqnarray}
e^{-2i\pi (J_2\alpha_{21}+J_3\alpha_{31})}\prod_{k=1}^{J_2+J_3}
\frac{u_{1,k}+\ialf}
{u_{1,k}-\ialf}
=1\ .
\label{mom_constr}
\end{eqnarray}
The  contribution of  a given  Bethe root solution to the energy is
\begin{eqnarray}
E=\sum_{k=1}^{J_2+J_3}\epsilon_k=\frac{\lambda}{8\pi^2}\,\sum_{k=1}^{J_2+J_3}
\frac{1}{u^2_{1,k}+\textstyle{\frac{1}{4}}}\ .
\label{energy}
\end{eqnarray}
%Here $L$ denotes the total length of the chain, which is just
%the total angular momentum, that is the sum of the eigenvalues
%$J_{1,2,3}$ of the three Cartan generators
%$H_{1,2,3}=\sum_k h_{1,2,3}(k)$.

%\newpage

\subsection{ Ground states of the  spin chain Hamiltonian}

%%%%%%%%%%%%%%%%%%%%%%%%%%%%%%%%%%%%%%%%%%%%%%%%%%%%%

The vacua  of the spin chain Hamiltonian  should
correspond to the ``BPS''  states of gauge theory that
 have zero anomalous dimensions.
For the  supersymmetric
deformation with  $\g_1=\g_2=\g_3=\g$  there are
at least two ways of finding them: (i) finding the generators of the
chiral ring and (ii) directly finding the solutions of the Bethe
equations which have zero energy.
Since for  general
nonsupersymmetric deformations the first option is not available,
we shall
therefore concentrate on the second  approach
(in the
supersymmetric  limit   we shall be able to compare the results with those of the
chiral ring analysis).

First,  let us note  that  since the dilatation operator is positive
semidefinite, the contribution of any Bethe root distribution
to the energy must be non-negative.
 Indeed, from the Bethe equations
(\ref{mainBetheSU3})-(\ref{auxBetheSU3}) one can  see
that, as in the undeformed case, the Bethe roots occur in complex
conjugate pairs which give  positive contributions to the energy
(\ref{energy}). Thus, the vacua of the spin chain fall into the
 two categories:

1) $J_2+J_3 = 0$ case in which  the energy  (\ref{energy}) is obviously zero;

2) configurations of Bethe roots for which $\epsilon_k=0$ for all $
k=1, \dots , J_2+J_3.$

\noindent
The first class corresponds to the classical vacuum of the spin chain
($\Tr\ \P_1^{J_1}$ operator)
which was  chosen  to derive the Bethe equations,
 as well as to its obvious $\IZ_3$
images.

For the  second class the expression \rf{energy}
 clearly implies that all rapidities $u_{1,k}$ are to be
infinite. This is similar to the case of other BPS states $\Tr ( \P_1^{J_1}
\P_2^{J_2} \P_3^{J_3} )_{symm}$
 in the undeformed $\N=4$ SYM
theory. As in the undeformed  case,  we will  take the difference
between any
two unequal  rapidities to also go to infinity. This is necessary
in order to focus on solutions which exist regardless
of whether $J_2+J_3$ is even or odd.
Unlike the undeformed   case,  however, due to the presence of the deformation
parameters $\g_i$ or $\alpha_{mn}$, not any such rapidity configuration  will be  a
solution of the Bethe equations and the cyclicity condition.

The cyclicity condition \rf{mom_constr} implies that the angular momenta $J_2$ and
$J_3$ must be chosen such that
\begin{eqnarray}
J_2\gamma_3-J_3\gamma_2=0\ .
\end{eqnarray}
Then, the main Bethe equation (\ref{mainBetheSU3}) further implies
that
\begin{eqnarray}
J_1\gamma_3-J_3\gamma_1=0\ .
\end{eqnarray}
Finally, taking the product of the auxiliary Bethe equations
(\ref{auxBetheSU3}) in the limit in which $u_{1,k}\rightarrow\infty$
implies that the third combination of the deformation parameters and the angular
momenta must vanish as well:
\begin{eqnarray}
J_1\gamma_2-J_2\gamma_1=0~~.
\end{eqnarray}
This discussion however is insufficient because it implies a rather
large degeneracy due to the fact that the auxiliary Bethe equation
(\ref{auxBetheSU3}) is nontrivial. It is,  however,  easy to see that, if
we focus on solutions which exist regardless of the parity properties
of $J_3$, we must have $|u_{2,k}-u_{2,l}|$ for all
$k\ne l$ (and therefore $u_{2,k}$ for any $k$) approach  the  infinity.

We  conclude that  for the general three real deformation parameters
$\g_i$ the spin chain Hamiltonian has three vacua corresponding to the
operators $
\Tr \  \Phi_i^J $, $i=1,2,3$,
as well as the  fourth vacuum corresponding to an operator containing
$J_i$ copies of $\Phi_i$ with $i=1,2,3$  provided
\begin{eqnarray}\la{co}
\epsilon_{ijk}J_j\g_k=0 \ , \  \ \  \ {\rm i.e.} \ \ \ \  J_i \sim \g_i \ .
\label{parallel}
\end{eqnarray}
Since $J_i$ are integer, this  forth vacuum can exist only for special
values of $\g_i$.
This matches the result of the string theory analysis in section 2.2.

In the case of supersymmetric deformation $\g_i=\g$ the above
condition \rf{co} becomes $J_1=J_2=J_3$. The existence of such $(J,J,J)$ BPS state
can be derived from the construction of the chiral
ring. The
 argument is the same as originally given for rational  deformation parameter  in
 \cite{BELE}, see also \ci{LM}.   Indeed, the $F$-term constraints
\begin{eqnarray}
e^{i\pi\beta}\Phi_1\Phi_2-e^{-i\pi\beta}\Phi_2\Phi_1=0\  ,
~~
e^{i\pi\beta}\Phi_2\Phi_3-e^{-i\pi\beta}\Phi_3\Phi_2=0\  ,
~~
e^{i\pi\beta}\Phi_3\Phi_1-e^{-i\pi\beta}\Phi_1\Phi_3=0
\end{eqnarray}
imply that, in the chiral ring, any single-trace operator can be
brought to the form
\begin{eqnarray}
\Tr(\Phi_1^{J_1}\Phi_2^{J_2}\Phi_3^{J_3})~~.
\end{eqnarray}
Then, the same  $F$-term constraints allow one to move any of the
$\Phi_i$ fields around the trace. In general, this multiplies the
initial operator by a phase whose argument is proportional to the
difference
between the number of fields of different types than the one which is
transported around the trace. For the operator to be an element of
the chiral ring it is necessary that this phase is unity which in turn
implies that
\begin{eqnarray}
J_1=J_2=J_3~~.
\end{eqnarray}

For rational $\g$ there are also  additional BPS states \ci{LM}
corresponding to rotating circular strings; they are, in fact, images of
certain BPS states in undeformed theory under the TsT
transformation. We have described them explicitly in
section 3.2. They are visible also in the Bethe ansatz. Indeed, for
rational $\gamma_i$ it is possible that
\begin{eqnarray}
\epsilon_{ijk}J_j\gamma_k\in\IZ~~.
\end{eqnarray}
This is enough to eliminate completely the deformation from the Bethe
equations. Thus, the the energy of the states with such
$(J_1,J_2,J_3)$ quantum numbers are identical to those in the
undeformed theory,
%NEW
i.e. they should be exact BPS states
(despite the theory not being supersymmetric for unequal $\g_i$).
%In particular, the completely symmetric combination of
%$(J_1,J_2,J_3)$ is a BPS operator ???
%{\bf please  rephrase this}.
\bigskip

We shall  discuss fluctuations  near these vacua as
 implied by the Bethe ansatz
equations  in Appendices A and B.

\bigskip
%%%%%%%%%%%%%%%%%%%%%%%%%%%%%%%%%%%%%%%%%%
\subsection{Comment on $U(N)$ vs. $SU(N)$ gauge theory}

It is worth emphasizing that 
the planar dilatation operator (\ref{hi}) and the 
corresponding
Bethe equations (\ref{mainBetheSU3})-(\ref{auxBetheSU3}) hold for the
$\beta_i$-deformation of the  ${\cal N}=4$ SYM
with $U(N)$ gauge group. The distinction between the $U(N)$ and 
 the $SU(N)$  case  is nontrivial 
 here even in the large $N$ limit.
 %AT
 More precisely, it is immaterial for ``long'' 
 single-trace operators we discuss in the main part of this paper 
 but matters for some ``short'' operators. 
  Indeed,  
  in the presence of the deformation, the $U(1)$ factor no
longer automatically decouples,  and that has interesting
 consequences; in particular, the $U(N)$ theory is not 
 automatically conformal  \cite{guf}, having running 
 couplings of $U(1)$ matter fields.
 
  It was recently
observed in  \cite{guf} and \cite{zan} that, while in the
supersymmetric $U(N)$ $\beta$-deformed SYM theory  the  
operators  $\Tr(\Phi_i\Phi_j)$ ($i \ne j$) 
 have
 nonvanishing one-loop anomalous 
dimension, their anomalous dimension is zero in the supersymmetric 
 $\beta$-deformed theory with $SU(N)$ gauge group. 
 The nonzero (in the large $N$  limit) 
 contribution in the $U(N)$  case comes
entirely from the non-decoupled $U(1)$ factor.

It is easy to see that the expression  \rf{hi} for the spin chain
Hamiltonian representing planar 1-loop $U(N)$ dilatation operator  
implies that the anomalous dimension of $\Tr(\Phi_i\Phi_j)$ is
\begin{eqnarray}
\Delta{\scriptstyle
(\Tr(\Phi_i\Phi_j))}=\frac{\lambda}{2\pi^2}\sin^2\pi\alpha_{ij} ~~,
\label{UNan}
\end{eqnarray}
where $\a_{ij}$ is given by \rf{alp}.
In the supersymmetric limit of equal deformation parameters 
this reproduces the result of \cite{guf} for
the $U(N)$ theory. 
The same result \rf{UNan} may be obtained from the Bethe
equations (\ref{mainBetheSU3})-(\ref{auxBetheSU3}) 
(and $\IZ_3$ symmetry).
 In the case of a single
excitation above the $(2,0,0)$ vacuum 
the Bethe equations simplify considerably; since the result
is determined by a single rapidity, it may,  in fact,
 be obtained from
the cyclicity condition which  trivially leads to
\begin{eqnarray}
\Delta{\scriptstyle
(\Tr(\Phi_1\Phi_j))}=\frac{\lambda}{2\pi^2}
\sin^2\pi \alpha_{1j} ~~.
\label{UNanB}
\end{eqnarray}
%where the index $k$ is not summed over. 
The other $\Tr(\Phi_i\Phi_j)$ anomalous dimensions
may be obtained using 
 $\IZ_3$ transformations or by changing the
vacuum of the spin chain. The expression 
 \rf{UNanB} is,  in fact, 
the  $L\rightarrow 2$ limit of the anomalous dimension of 
$\Tr(\Phi_1^{L-1}\Phi_j)$.

In the case of the deformation of the $SU(N)$ SYM
theory these  anomalous
dimensions vanish due to an ``accidental cancellation''.
As was pointed out in \cite{guf} in the case of the supersymmetric 
  deformation of the $SU(N)$ theory, 
  the superpotential
contribution to the potential can be written as
\begin{eqnarray}\la{exep}
V=\sum_{a=1}^{N^2-1} \left[
  | \Tr([\Phi_1, \Phi_2]_{_\b} T^a)  |^2
+ | \Tr([\Phi_2, \Phi_3]_{_\b} T^a)  |^2
+ | \Tr([\Phi_3, \Phi_1]_{_\b} T^a)  |^2  \right] \ ,
\end{eqnarray}
where $[\Phi_1, \Phi_2]_{_\b}= \Phi_1 * \Phi_2 - \Phi_2 * \Phi_1=
e^{  i \pi \b}\Phi_1 \Phi_2- e^{ -  i \pi \b}
 \Phi_2 \Phi_1$ 
 and $T^a$ are the $SU(N)$ generators.
As a result,  the anomalous dimension of
holomorphic 2-field operators is proportional to
 $\Tr\ T^a$ which
vanishes. Clearly, such a cancellation does not occur in the 
deformed $U(N)$ theory. 
In the undeformed 
$U(N)$ ${\cal N}=4$ SYM theory, 
the dilatation operator \rf{hah},\rf{haoo}  combined with 
the cyclicity of the trace still leads to the  vanishing anomalous
dimension for  $\Tr(\Phi_i\Phi_j)$.
%AT
\foot{The reason why 
there is a  difference between  the $SU(N)$  and $U(N)$ cases 
even  in the large $N$ limit  has to do  with non-decoupling 
of $U(1)$ part of scalar multiplets (in the pure gauge field sector 
$U(1)$ part of $U(N)$ always decouples at large $N$); it 
is also special to the case of length-2 operators.
A quick way to see  why the double-trace quartic scalar 
vertex present in the $SU(N)$
case  ($(T^a)^i_j (T^a)^k_l = \delta^i_l \delta^k_j  - 
{ 1 \ov N} \delta^i_j \delta^k_l $) 
  does contribute  to the anomalous dimension  of the 
Tr$(\P_i \P_j)$  operator in the same way as the single-trace vertex 
is to consider the generating functional $Z(k)$ for the correlators 
of the Tr$(\P^2)$ operators (suppressing all indices).
The corresponding action will look like 
$S= { N \ov \l} \int [ ... + \Tr (\P^2)^2 +
 {1 \ov N} ( 1 + k(x) )  \Tr (\P^2) ] $, 
 where  we do not make distinction between 
 the structure of $\P^2$ term in the vertex  and in the operators
 for which $k(x)$ is a source.  Then it is clear that derivatives over
 $k(x)$ will scale as $N^0$, which is the same 
 cylinder-diagram scaling as for the 
 2-point function  of $\Tr (\P^2)$  with insertions of 
 $\Tr \P^4$  vertex. We thank K. Zarembo  for 
 a discussion of this  point.}

It is not hard to see that, in the case of the nonsupersymmetric
$SU(N)$ gauge 
theory with unequal $\b_i= \g_i$ the 
cancellation due to the tracelessness of gauge group generators
also takes place.
As was mentioned above,
 the nonsupersymmetric $\g_i$-deformed
  theory is obtained by replacing in
the  component Lagrangian of the ${\cal N}=4$ SYM
the ordinary product of fields with the noncommutative product
\begin{eqnarray}
\Phi_i * \Phi_j(x)~~\mapsto~~\lim_{y\rightarrow x}
e^{-i\pi \alpha_{mn}\hh_m(x)\hh_n(y)}\Phi_i(x)\Phi_j(y)~~,
\label{starp}
\end{eqnarray}
where  again  $\alpha_{mn}= -\epsilon_{mnk}\g_k$, 
 $\hh_n$ are the three global $U(1)$ symmetry generators 
(i.e. $\hh_n \P_i = \delta_{ni} \P_i$), 
 and summation over $m,n$ is assumed.
Thus the potential relevant for  
the calculation of anomalous dimensions
of scalar operators  may be written as 
\begin{eqnarray}
V=\sum_{a=1}^{N^2-1}\left[
  | \Tr([\Phi_1, \Phi_2]_{_{\g_3}} T^a)  |^2
+ | \Tr([\Phi_2, \Phi_3]_{_{\g_1}} T^a)  |^2
+ | \Tr([\Phi_3, \Phi_1]_{_{\g_2}} T^a)  |^2 \right] \ .
\end{eqnarray}
Using this form of $V$ for the calculation of
anomalous dimensions of  the same $\Tr(\Phi_i\Phi_j)$ operators
it is easy to
see  that their  anomalous dimensions are  proportional to
\begin{eqnarray}
 ( e^{2 i\pi \g_k\eps_{ijk}} - e^{-2 i\pi\g_k\eps_{ijk}}) \ \Tr\ T^a \
\end{eqnarray}
and thus vanish again in the $SU(N)$ gauge group case
regardless the values of
the  deformation parameters $\g_i$.

Such cancellations appear   not to exist for longer operators,
even in the supersymmetric $\g_i=\g$ case. Indeed, in
that case the chiral ring argument appears to imply that the 
only
chiral operators are
 in the representations $(\JJ,0,0)$, $(0,\JJ,0)$,
$(0,0,\JJ)$ and $(\JJ,\JJ,\JJ)$.

Given that it is possible to break supersymmetry by an 
arbitrarily  small amount (the deformation parameters $\g_i$ are
continuous) and that the spectrum of 1-loop anomalous dimensions
has a gap, it is reasonable to search for protected operators in the
nonsupersymmetric theory among the protected operators in its
supersymmetric limiting case. %As we have seen throughout the paper,
Since we have argued that we already know
all such operators with  vanishing anomalous
 dimensions,  we do not expect
  additional operators with vanishing
anomalous dimensions for sufficiently 
small $\g_1,\,\g_2,\,\g_3$.

\bigskip

%%%%
%% Changes below ----------
%%%%

An interesting question is what is the dual string theory
 prediction for
the anomalous dimensions of operators $\Tr(\Phi_i\Phi_j)$ or,
alternatively, which theory does the dual string theory describe: 
the $U(N)$ or the $SU(N)$ gauge theory.\foot{
There are
 several notable differences between the $\g_i$-deformed
theory and ``standard''  noncommutative 
field theories where one finds  non-decoupling of the $U(1)$ 
gauge  fields. 
As was mentioned above, 
the *-product describing the $\g_i$ deformation can be thought of as a
Moyal product based on the Cartan generators of the
remnant of the  ${\cal N}=4$ SYM 
R-symmetry group. It  acts nontrivially 
only on the fields carrying non-zero  charges under these generators,
i.e. the fermions and the scalar fields, but {\it not} the gauge fields.
%%This implies that, from a field theory standpoint, one is allowed to
Therefore, one is allowed 
to truncate away the $U(1)$ gauge field, but not the $U(1)$ scalars and
fermions. An analog of this non-decoupling 
in matter sector  may be observed on the string theory side
as well. Realizing the gauge theory on a collection of coincident
D3-branes, the decoupling of the diagonal $U(1)$ degrees of freedom is
associated to  the translational invariance of the collection of
branes. The form of the string background 
 \rf{me}-\rf{angl} dual to the deformed gauge theory
suggests  that in a similar set-up 
%% , the same setup requires
 there should  exist nontrivial fluxes in the space 
transverse to the branes. These fluxes  will 
 break translational invariance and lead to the non-decoupling of
the fields carrying charges under the flat space ``transverse'', i.e.
internal,  symmetry
group. Translational invariance if,  however, 
 maintained along the branes
and thus the gauge fields continue to decouple.
%%AT
This is also 
reflected in the fact that here we have the standard $AdS_5$ factor
 in the  geometry 
while in the noncommutative case the  solution of \ci{rum} does 
not have an  $AdS_5$ asymptotics.}
The answer 
is nontrivial here and appears to be $SU(N)$: 
 even though the diagonal $U(1)$ gauge fields still decouple
in the presence of the deformation, the $U(1)$ scalars 
and fermions do not. Due to the absence of coupling with the gauge
fields, it is expected that the RG beta-functions of the
 couplings of
these $U(1)$ fields is positive and they flow to 
zero at low energies.\foot{We thank O.~Aharony, J.~Maldacena 
and E.~Sokachev for
discussions on this point.}
 For the supersymmetric deformation this is indeed the case so only 
 the $SU(N)$ theory  is conformal
%scenario is indeed realized
 \cite{dor,guf}. 
%%
%% R
%%
For the nonsupersymmetric deformation new effects may
appear. For example, the three terms 
$|[\Phi_i,\,\Phi_j]_{\beta_k}|^2$ are related by a $\IZ_3$ symmetry
which also acts on the deformation parameters. Thus, it is possible 
that their coefficients undergo different renormalization and, while
equal at one-loop level, at higher loops
they may have different values at the fixed point. Similarly to the
case of the supersymmetric deformation, further new
phenomena may appear for rational values of $\beta_i$ related to the
appearence of additional operators with vanishing anomalous dimensions.
  
The observation that the coupling of the $U(1)$ matter fields runs 
suggests that the string theory in the deformed background 
\rf{me}-\rf{angl} describes the deformation of the conformal 
$SU(N)$ gauge
theory. The anomalous dimensions of holomorphic operators
$\Tr(\Phi_i\Phi_j)$ with $i\ne j$ were  computed to two loops in 
the supersymmetric case  in 
\cite{zan} and found to be subleading in the $1/N$ expansion. It may
be possible that these operators remain marginal to all orders in the
planar limit. It would be interesting to check this explicitly, by
analyzing in supergravity the corrections to the masses of the 
fields in the ${\bf 20}$ of $SO(6)$ once the
 deformation is turned on.

\bigskip

\renewcommand{\theequation}{5.\arabic{equation}}
 \setcounter{equation}{0}

%%%%%%%%%%%%%%%%%%%%%%%%%%%%%%%%%%%%%%%%%%%%%%%%%%%%%%%%%%%%%%%%%%%%%%
\section{Coherent state effective action}
%%%%%%%%%%%%%%%%%%%%%%%%%%%%%%%%%%%%%%%%%%%%%%%%%%%%%%%%%%%

In the case of undeformed  ${\cal N}=4$ SYM -- \adss string duality
the  matching  of predictions for energies of states with large
quantum numbers
can be done in a universal way by comparing the effective
action for the long wave length spin chain excitations
with the
effective action for the ``slow'' world sheet modes  obtained as a
limit of the classical string action
after  separating
the ``fast'' collective  string modes  \ci{KRUC}.

The relevant spin chain degrees of freedom can be described
by the spin coherent states $|n\rrangle$  with the action
\be
\la{acc}
S_{ coh}= i\llangle n|\partial_t|n\rrangle-\llangle n|{H}|n\rrangle\ ,
\ee
which appears in the exponent in the coherent state path integral.
The limit  one is  interested in, i.e. $\JJ\to \infty, \
  \tl = {\l\ov \JJ^2}, $
is the semiclassical limit for the spin chain path integral
in which one can take the continuum limit keeping only the leading
2-derivative terms in $S_{coh}$.

In the case of the 2-spin sector, i.e. operators  $\Tr ( \P_1^{J_1}
 \P_2^{J_2}+ ...)$,
 this
strategy was successfully applied to the supersymmetric deformed theory
\cite{frt}, demonstrating  the equivalence  of the two  effective actions.
The case of the 3-spin sector (and larger non-holomorphic sectors) is,  however,
somewhat different.
%A hint of potential complications appeared in the above
% discussion of  fluctuations around the $(\JJ,0,0)$ vacuum,
%where the intuition on the long wave length modes coming from the
%undeformed theory led to inconsistent equations.
In the context of the
coherent state continuum limit the  problem arises in that
a  naive derivation that follows the same strategy as in the undeformed 3-spin
case \ci{hlo,st} or the deformed 2-spin case \ci{frt}
 leads to an  effective
action that does not  not properly describe all expected vacuum states
as seen in the Bethe  ansatz  and also on the string theory side.

Indeed,
as in the case of the 2-spin sector in \ci{frt},
the fact that the ``Wess-Zumino''
term in the string action \rf{eexx} is independent
 of the deformation parameters suggests that we may
use the same coherent state as in the $su(3)$ sector
in the undeformed theory.
Then the resulting  effective action
 as found in the continuum limit from \rf{acc} with $H$ given by \rf{hah},\rf{hp}
 happens to contain the potential
\begin{eqnarray}
V_{\it naive}=\g_1^2 \r_2^2\r_3^2+\g_2^2 \r_3^2\r_1^2+\g_3^2 \r_1^2\r_2^2\ , \ \ \ \ \
\ \   \sumi \r^2_i =1 \ ,
\label{na}
\end{eqnarray}
where we used the same notation $\r_i$ as  in  the string-theory
potential \rf{poo}
for the corresponding coherent state parameter. Compared to \rf{poo}
this
potential  misses the last $\r^6$ term. As a result,
it
captures the vacua $(\JJ,0,0)$, etc., but  misses the non-trivial one
with  $(J_1,J_2,J_3)\sim (\g_1,\g_2,\g_3)$
or $(J,J,J)$ in the equal $\g_i$ case.

To understand the source of the problem it is useful
to recall the story of BPS vacua in the $su(3)$
sector of the undeformed theory. There the spin chain Hamiltonian
\rf{haoo} containing permutation operator
has vacua represented by all  totally symmetrized
products of the three chiral fields  $\Tr (\ppp)_{symm}$.
Apart from  $\Tr\  \P_1^{J}$, $\Tr\ \P_2^{J}$ and $\Tr\   \P_3^{J} $, i.e.
$(\JJ,0,0)$, $(0,\JJ,0)$ and $(0,0,\JJ)$  vacua these are {\it
not} ``slow'' modes
of the spin chain: the field component $\P_i$
in general changes rapidly from site to site.
The coherent state operators  that are mapped onto semiclassical
string  states (in this case geodesics or point-like strings  all of
 which here are BPS)  are particular linear combinations of these quantum
spin chain  vacua:\foot{All quantum BPS vacua correspond to Kaluza-Klein
modes (spherical harmonics), while their particular  coherent
combinations have  semiclassical interpretation
as point-particles moving along geodesics  of $S^5$.
All such geodesics are related by $SO(6)$ rotations.}
 if
 the generic  coherent state operator is
$\Tr (\prod^\JJ_{k=1}  [ \sumi n_i(k)\P_i ])$ then the vacua  correspond to constant
$n_i$, i.e. to $\Tr ( \sumi n_i \P_i )^\JJ$,
which are indeed  linear combinations of symmetrized products.

In the present deformed  case we do not have all possible $(J_1,J_2,J_3)$
BPS quantum vacua to built a coherent linear superposition, and,
moreover, the nontrivial vacuum $(J_1,J_2,J_3)\sim (\g_1,\g_2,\g_3)$
 is not a ``slow''  state. Yet, the fact that it is naturally found also  on
 the string theory side suggests that there  should be a way to capture it
in the coherent state action.

The source of the problem  thus  appears to be in the choice of a
description of the relevant spin chain modes by coherent states.
One is either to generalize  the definition of coherent states,
or, alternatively,  to  use the ``undeformed''  coherent states but
choose a different representative in the class of equivalent
spin chain Hamiltonians with  the same spectrum.

The latter option is equivalent to  changing the basis.
The spin chain  Hamiltonian
represents the gauge-theory anomalous dimension matrix in the basis of
single-trace single-term operators.
As we shall show below, there is
a way to choose   a more suitable basis
 basis so that the resulting coherent state action
\rf{acc} adequately  describes the ``low-energy'' approximation with
all vacua included, and, moreover, matches its string-theory
counterpart \rf{eexx}.

\bigskip

%%%%%%%%%%%%%%%%%%%%%%%%%%%%%%%%%%%%%%%%%%%%%%%%%%%%%%%%%%%%%%%%%%%
We shall start with reviewing the choice of the coherent states which will
be the same as in the undeformed $su(3)$ case. We shall  then
describe a change of basis leading
 to an equivalent (but more appropriate for
the   low-energy  description with standard set of coherent states)
Hamiltonian $\td H=  U^{-1} H U$.
Finally, we shall use $\td H$ to
compute  $ S_{coh}$ in \rf{acc}  and
find its continuum limit.
%%%%%%%%%%%%%%%%%%%%%%%%%%%%%%%%%%%%%%%%%%%%%%%%%

%%%%%%%%%%%%%%%%%%%%%%%%%%%%%%%%%%%%%%
\subsection{The coherent state}
%%%%%%%%%%%%%%%%%%%%%%%%%%%%%%%%%%%%%%%%%%

While the standard  definition of coherent states based on
global symmetry of the Hamiltonian\foot{In general, it is given by an $G/G_0$
transformation applied to a ground state, where $G$ is a
 symmetry of $H$ and $G_0$
 is a  symmetry of the ground state.}
does not formally apply in the present
deformed case, we can still
use the  $SU(3)$-invariant
 coherent state  which is  a tensor product
 over the spin 1 (3-component)
chain sites of a state obtained by a $3 \times 3$ rotation which keeps fixed
some specified 3-vector:
\begin{eqnarray}
R=R(h)R(k)\ ,
~~~~
R(h)&=&{\rm diag}\,(e^{ih_1},\,e^{ih_2},\,e^{ih_3})\ , ~~~~\sumi h_i=0\ , \cr
R(k)&=&\id-2|k\rangle\langle k|\ , ~~~~~~~\langle k|k\rangle =1\ .
\end{eqnarray}
This state can be parametrized by an
element of  $\IC\IP^2$.
With appropriate redefinitions
($
k_1=\frac{1}{2}\sqrt{1-n_1} , \
k_2=-\frac{n_2}{2\sqrt{1-n_1}},\
k_3=-\frac{n_3}{2\sqrt{1-n_1}},
$)
the   coherent state can be written as:
\begin{eqnarray}
|n\rangle = n_1 |1\rangle+n_2 |2\rangle+n_3 |3\rangle\ ,
\end{eqnarray}
where
$e^m_n$ in \rf{yy} acts on $|i\rangle$ as
\be
e_i^m|j\rangle= \delta_{ij}  | m \rangle  \ ,  \ee
and
\begin{eqnarray}
n_i   = m_i e^{i h_i}\ , \ \ \ \ \ \ \ \ \ \ \ \
\sum_{i=1}^3m_i^2=1\ ,
~~~~~
\sum_{i=1}^3 h_i=0 \ .
\end{eqnarray}
On the string theory side (cf. \rf{eexx})
$m_i$ will correspond to $\r_i$, and
$h_i$ to $\p_i$ but for generality  we shall use
 this separate  notation.

The total spin-chain state  coherent state is then
\begin{eqnarray}
|n\rrangle=|n\rangle_1\otimes |n\rangle_2\otimes \dots
\otimes|n\rangle_L\ , \ \ \ \ \ \ \
|n\rangle_k= \sumi n_i(k) |i \rangle \ . \la{cohi}
\end{eqnarray}
It thus corresponds to the operator
\begin{eqnarray}
\Tr[  (\sumi n_i(1)\Phi_i)  ... (\sum^3_{j=1} n_j(L)\Phi_j) ] \ ,
\end{eqnarray}
up to the cyclicity of the trace composed with
cyclic permutations of the site labels $1,2,\dots, L$.\foot{
One may also include the factor of
${1}\ov {\sqrt{L}}$ that  makes the state (and the operator) unit normalized.
This factor ends up playing no role; it  cancels because there
are always $L$ identical terms contributing to the expectation value
of any operator.}

With this choice of the coherent state the first WZ term in the
continuum effective action
\rf{acc} has the same standard form as in the undeformed case:
\begin{eqnarray}\la{wz}
S_{WZ}=\JJ \int^{2\pi}_0 { d\sigma \ov 2 \pi} \
\sumi  m_i^2{\dot h}_i \ .
\end{eqnarray}

%%%%%%%%%%%%%%%%%%%%
\subsection{The change of basis:
 choice of an equivalent Hamiltonian}
%%%%%%%%%%%%%%%%

Let us first recall the meaning of the change of basis for the spin
chain Hamiltonian.
 The precise statement about  the relation between
string energies and the gauge-theory
anomalous dimensions is that the string energies
are equal to the eigenvalues of the dilatation operator.
 The latter are
computed by finding the anomalous dimension matrix and then
diagonalizing it. The spin chain Hamiltonian is the
anomalous dimension matrix, that is  the dilatation operator in
the basis of operators used to compute the anomalous dimension matrix,
\be
\Delta O_A = H_A^B O_B ~~  ,
\label{anomdim}
\ee
where $A,B$ are  multi-indices.
The spin chain Hamiltonian \rf{hah},\rf{hi},\rf{hp}
 was computed in the ``standard"
basis, that is the basis of single-term single-trace operators
$\Tr(\Phi_{i_1}...\Phi_{i_L})$.
Changing this basis leads to a change of the expression for  the spin chain
Hamiltonian. {}{}From (\ref{anomdim}) we see that a general change
of basis $O_A=U_A^B{\tilde O}_B$ acts on the Hamiltonian
by  the transformation
\be \la{pu}
%\Gamma \rightarrow {\tilde \Ggamma}= U^{-1} \Gamma U
    H   \rightarrow {\widetilde H} =     U^{-1} H U \ .
\ee
Since the original $H$ in \rf{hp} contains only nearest-neighbor
interactions
it is  clear that the operator $U$  which we need  should be
nontrivial since it should be able to
generate in the continuum limit of $S_{coh}$ higher than 4-th
powers of the ``radii'' $m_i$ in order to get an  effective potential
that will have more than just three obvious vacua $m_i=(1,0,0)$, etc.

Consequently,  $U$ cannot be a site-wise tensor product.
It  is natural to try the next simple possibility,
 i.e. a product of
operators overlapping only on  one site
\begin{eqnarray}
U=\prod_{k=1}^L ~U_{k,k+1}~~,
\label{factorize}
\end{eqnarray}
with the  additional assumption that $U_{k-1,k}$ commutes with
$U_{k,k+1}$. Without this additional assumption $U^{-1}HU$ would be a
double sum over the spin chain sites and,  therefore,
would lead to   a nonlocal
effective action.
Combined with  the observation that the original spin chain
Hamiltonian  can be written in the form \rf{vv}, i.e.
\begin{eqnarray}\la{jj}
~H = \sum_{k=1}^LH_{k,k+1}=\sum_{k=1}^L
\cU_{k,k+1}^{-1}H^{(0)}_{k,k+1}\cU_{k,k+1}\ ,
%\cr &&H^{(0)}_{k,k+1}
%=\frac{\lambda}{8\pi^2}\left[
%\id_k\otimes\id_{k+1}-{\cal P}_{k,k+1}
%\right]~~~~~~ O_{k,k+1}=e^{i\alpha_{mn}e_m^m(k)e_n^n(k+1)}
\end{eqnarray}
this suggest the following natural  ansatz:
\be
U_{k,k+1}(\xi)=(\cU_{k,k+1})^\xi =e^{i\xi\pi \sumn
\alpha_{mn}e_m^m(k)e_n^n(k+1)}
= \sum_{m,n=1}^3 e^{ i\xi \pi\alpha_{mn}  }   e_m^m(k)e_n^n(k+1) \ ,
\label{transf}
\ee
\be
U^{-1}_{k,k+1}(\xi)=U_{k,k+1}(-\xi)\ , \ \ \ \ \ \
 U_{k,k+1}(1) = \cU_{k,k+1}  \ .
\ee
Here $\alpha_{mn} $ are the same phases as  in \rf{hp} and $\xi$ is
a parameter.
This ansatz is, in fact,  quite unique.
For example, allowing for off-diagonal generators in the exponent would
violate the locality requirement.\foot{For generic complex 
$\xi$ the transformation
 above  acts on the basis
operators by adding a phase and a rescaling depending 
on the order of fields in a monomial, so 
this is a rather simple  change of the basis.}

It is relatively easy to find the transformed Hamiltonian. The main
observation is that in each term in the sum defining $\widetilde H$
in \rf{jj} all factors
of $U$ cancel out except for those which have nontrivial overlap with
$H_{k,k+1}$. We find
\begin{eqnarray}
{\widetilde H}&=&U^{-1}(\xi) HU(\xi) \cr
&=& \sum_{k=1}^L
U_{k-1,k}(-\xi)U_{k,k+1}(-\xi+1)U_{k+1,k+2}(-\xi)H^{(0)}_{k,k+1}
U_{k-1,k}(\xi)U_{k,k+1}(\xi-1)U_{k+1,k+2}(\xi)\cr
&\equiv& \sum_{k=1}^L {\widetilde H}_{[k]}\ ,
\end{eqnarray}
where $[k]$ is used to  indicate the dependence
on the sites  $ k-1,k,k+1,k+2$.
Using trivial identities following from the properties of $e_m^n$,
${\widetilde H}_{[k]}$ can be simplified to:
\begin{eqnarray}
{\widetilde H}_{[k]}&=& \frac{\lambda}{8\pi^2} {\widetilde \H}_{[k]}=
\frac{\lambda}{8\pi^2} \sum_{m,n,p,r,q,t=1}^3
e^{i\pi\xi(\alpha_{mr}-\alpha_{mn})}
e^{i\pi\xi (\alpha_{qt}-\alpha_{pt})}
e^{i\pi(\xi-1)(\alpha_{rq}-\alpha_{np})}
\cr
&&~~~
\times
e_m^m(k-1) \bigg[ e_n^n(k)  e_p^p(k+1)
H^{(0)}_{k,k+1}e_r^r(k)  e_q^q(k+1)\bigg]  e_t^t(k+2)~~.
\end{eqnarray}

%%%%%%%%%%%%%%%%%%%%%%%%%%%%%%%%%%%%%%%%%%%
\subsection{Continuum limit and the effective action}
%%%%%%%%%%%%%%%%%%%%%%%%%%%%%%%%%%%%%%%%%%%%

The important piece of information in constructing the effective action
is  the cyclicity property of the states described by it. In the
initial form  (\ref{jj}) of $H$ the states the Hamiltonian acted on
were periodic. An arbitrary change of the basis may affect this and lead
to non-periodic states. The
transformation (\ref{transf}) has the crucial  property that it
commutes with the shift operator. Therefore, the states the
transformed  Hamiltonian acts on continue to be cyclically symmetric.
This implies that we are allowed to use the coherent state
(\ref{cohi}) to construct the effective action.

Using the expression \rf{haoo},\rf{yy} for the
undeformed Hamiltonian in the $su(3)$ sector or
$
[H^{(0)}_{k,k+1}]{}_{rq}^{np}=
\delta_r^n\delta_q^p - \delta_r^p\delta_q^n$
it follows that the expectation value of ${\widetilde \H}_{[k]}$
in the above  coherent state $|n\rrangle$ \rf{cohi}
is:
\begin{eqnarray}
&&\llangle n| {\widetilde \H}_{[k]}
|n\rrangle
=\sum_{n,p=1}^3
\Big[
(m_n(k)){}^2 (m_p({k+1}) ){}^2 \nonumber \\&& - \
 \sum_{q=1}^3
 (m_q ({k-1})){}^2  e^{i\pi\xi(\alpha_{qp}-\alpha_{qn})} \nonumber \\
&&\times \
e^{-2i\pi(\xi-1)\alpha_{np}}{m_n (k)  m_p(k)  m_p({k+1})  m_n ({k+1}) }
e^{i (h_n(k) -h_n ({k+1}) -h_p(k)  + h_p ({k+1}) ) }   \\
&&\times  \sum_{r=1}^3
 (m_r({k+2}) ){}^2 e^{i\pi\xi(\alpha_{nr}-\alpha_{pr})}
\Big]\ . \la{som}
\nonumber
\end{eqnarray}
%where we used the notation $m_k^i\equiv m_i(k)$, $h^i_k\equiv  h_i(k)$.
Expanding this expression in $\alpha_{mn}$ (i.e. in the deformation
parameters  $\g_i = \frac{1}{2}\ep_{inm} \a_{mn}$)
  and in  the spin chain
spacing $a$ up to the second order and suitably combining the resulting
terms we find ($\del m(k) \equiv  { m({k+1}) - m(k) \ov a} $)
\begin{eqnarray}
\llangle n| {\widetilde \H}_{[k]}
|n\rrangle
&\simeq& \sumi (\partial m_i (k) )^2
+ \sum_{p<n=1}^3
[\partial h_p(k) -\partial h_n(k) + \frac{2\pi}{a}
\alpha_{pn}]^2(m_p(k)  m_n (k) )^2 \nonumber  \\
&&
\vphantom{\Big|}
-\ 2\xi(1-\xi)(\frac{2\pi}{a})^2 ( \sum_{p<n=1}^3 \alpha_{pn})^2
[m_1 (k)   m_2 (k)  m_3 (k) ]^2 \ .
\end{eqnarray}
As usual, the sum over sites is replaced by an integral over $\s \in [0,\,2\pi]$, and  using the
relation between the lattice spacing $a$ and the
length of the chain ($L \equiv \JJ = \sumi J_i$)
\begin{eqnarray}
a=\frac{2\pi}{\JJ} \
\end{eqnarray}
we get for  the
continuum limit of the coherent state expectation value
of the transformed  Hamiltonian (here $\tl = { \l \ov \JJ^2}$,\  $ '= \del_\s$)
\begin{eqnarray}
\llangle n| \widetilde H|n\rrangle=\ha \JJ \tl \int_0^{2\pi} {d\sigma\ov 2 \pi}
\Bigg[ &&
\left(m_1\, m'_2-m_2\, m'_1\right)^2
+(h'_1-  h'_2+ \JJ \alpha_{12})^2(m_1 m_2)^2\cr
&+&\vphantom{\Big|}
\left(m_2\, m'_3-m_3\, m'_2\right)^2
+( h'_2-  h'_3+ \JJ \alpha_{23})^2(m_2 m_3)^2\cr
&+&\vphantom{\Big|}
\left(m_3\,  m'_1-m_1\,m'_3\right)^2
+(  h'_3- h'_1+ \JJ \alpha_{31})^2(m_3 m_1)^2\cr
&-&\vphantom{\Big|}
2\xi(1-\xi)\left(\JJ\alpha_{12}+\JJ\alpha_{23}+ \JJ\alpha_{31}\right)^2
(m_1m_2m_3)^2~~~
\Bigg]
\label{vevHU}
\end{eqnarray}
As required in our scaling limit, the action is finite (modulo the overall
factor of $\JJ$)
 for
fixed $\tl = {\l \ov \JJ^2}$ and $\JJ \a_{mn}= - \ep_{mni} \JJ \g_i $.
It   thus describes a particular sector of low-energy
excitations of the spin chain
 (``macroscopic spin waves'') which correspond to semiclassical
fast-moving  strings  in the 3-spin sector.

We are now in position to determine the free parameter $\xi$ in the
definition of $U(\xi)$ by requiring that the effective action
reflects  the correct vacuum
structure. Intuitively, the existence of the vacua $(\JJ,0,0)$, $(0,\JJ,0)$
and $(0,0,\JJ)$ should  not impose any constraints on $\xi$ because $U$ acts
trivially on these states. This can indeed be verified by the explicit
calculation (the corresponding critical points are $m_1=1,m_2=0,m_3=0$, etc).
 The existence of the  $(J_1,J_2,J_3)$ vacua  with  $J_i\sim \g_i $
   does, however,
require the specific value of $\xi$.
For example, in the case of $\g_i=\g$  the corresponding critical point of the
potential  term  in  (\ref{vevHU}) is $m_i=\pm\frac{1}{\sqrt{3}}$
and the value of the potential at this point is
$ V = \ha \JJ \tl ( \JJ \g)^2  [\frac{1}{3}- \frac{2}{3}\xi(1-\xi) ] $.
By requiring that it vanishes  we get
\begin{eqnarray}
\xi=\frac{1}{2}(1\pm i)~, \ \ \ \ \ \ \ \ \ \ \ 
   2\xi(1-\xi) =1 \ .
\label{critval}
\end{eqnarray}
One can  directly verify that in the general $\g_i$ case
 the effective potential with
 these values of $\xi$  is non-negative
and  vanishes only at the required  four critical  points.

The full coherent state action which correctly
reproduces the spin chain vacuum structure is thus
given by  the difference of \rf{wz} and  \rf{vevHU}
with  the coefficient of the last term being $2\xi(1- \xi) =1$.
Remarkably, it then  also reproduces
the fast string action  \rf{lal},\rf{eexx}
with the identification  $\r_i=m_i, \ \p_i=h_i$.

Let us note that  the complex  value of $\xi$ implies  that the
transformation in \rf{factorize},\rf{transf}  is not  unitary. This manifests
itself at higher orders in the $\g_i$ expansion and therefore 
implies that at higher loops a further change of basis is necessary.
While the unitarity of the basis change is not a required condition,
we suspect that there may exist also a unitary change of basis that 
leads to the same real coherent state effective action.

\renewcommand{\theequation}{6.\arabic{equation}}
 \setcounter{equation}{0}

%%%%%%%%%%%%%%%%%%%%%%%%%%%%%%%%%%%%
\section{Concluding remarks}
%%%%%%%%%%%%%%%%%%%%%%%%%%%%%%%%%%%%%

In this paper we have studied an example of  large $N$ AdS/CFT  duality
in a non-supersymmetric context.

The string theory we considered
is obtained  from the \adss string theory by a combination of T-dualities and
shifts of angular coordinates  and is parametrized  in addition
to the radius $R = \l^{1/4} $  ($\a'=1$)
of the  $AdS_5$ space  by the three real parameters $\tg_i = R^2 \g_i$
which determine the shape of the deformed $S^5_{\g_i}$  space.
The  special case of equal $\g_i=\g$  corresponds to the
supersymmetric deformation of \adss string theory  introduced in \ci{LM}
and further studied in \ci{frt,f}.

The dual gauge theory has the same field content as the  $\N=4$ SYM
theory, but with scalar quartic interactions  and Yukawa couplings
being  ``*-deformed'' using $\g_i$ as phase multiplying the 
$U(1)$-charges of the fields \ci{f}. 
The three $U(1)$ symmetries and the corresponding 
charges 
are inherited from the $SU(4)$ R-symmetry
of ${\cal N}=4$ SYM theory.
%} Its global
%symmetry  is reduced by the deformation from $SO(6)$ to $U(1)^3$.
In the case of $\g_i=\g$  the gauge theory 
 becomes the
exactly marginal $\N=1$ supersymmetric deformation of  
the $\N=4$ SYM theory
with real deformation parameter $\b=\g$.

We have compared the energies of the semiclassical strings
in $AdS_5 \times S^5_{\g_i}$ geometry having  three large   angular momenta
in $S^5_{\g_i}$
to the 1-loop anomalous dimensions  of the corresponding gauge-theory
scalar  operators and found that they match  just as it was the case
in the $SU(3)$ sector of the standard  \adss  duality \ci{ft2,Mina,hlo,st}.

In particular, in the supersymmetric special case of $\g_i=\g$
this extends the result of \ci{frt} from the 2-spin sector to the 3-spin
sector. This  extension turns out to be quite nontrivial.
To match the corresponding low-energy effective actions  on the string
theory and the
gauge theory side one  is to make a special choice of the spin chain
Hamiltonian  representing the 1-loop gauge theory dilatation operator.
This  choice
      is ``adapted'' to the low-energy or semiclassical
       approximation, i.e. it
allows one to capture the right vacuum states and  the
``macroscopic spin wave''
sector of states of the spin chain in the continuum coherent state
effective action.

Our results suggest that  some quantitative aspects
of the AdS/CFT correspondence may be less sensitive to
the presence of supersymmetry than it was  previously expected.
%They also stress the utility of the approach based on the
%low-energy effective actions of Landau-Lifshits type.
There are, 
of course,  many ways to break supersymmetry of the original
maximally supersymmetric AdS/CFT set-up. The important observation of
\ci{LM} extended in \ci{f} to the non-supersymmetric case
is that the TsT  duality preserves the regularity of the geometry
and thus leads to tractable examples of the duality. Also,
the present theory  has continuous tunable parameters
 which is an
advantage over the  orbifold \ci{ide}
models.\foot{Other $\N=1$ or $\N=0$  models
based on replacing $S^5$ by  less-symmetric spaces
of different topology (like $T^{1,1}$ \ci{kw} or $S^2
\times S^3$)  and 
corresponding to  non-perturbative isolated conformal
fixed points on the gauge
theory side,  appear to be under less theoretical  control.}

One of the by-products of our investigation  of the
spectrum of fluctuations near nontrivial $(J_1,J_2,J_3)$ vacuum
of deformed theory on the gauge theory side is the discovery
of a new type of solutions of the Bethe equations for the
3-spin sector  of deformed theory (see Appendix A and B). Switching on the
deformation parameters lifts the degeneracy of the spectrum of conformal
dimensions of the $\N=4$ SYM theory and leads to new non-trivial
relations between  the structure of solutions of the Bethe equations
and  dimensions of gauge-theory operators with given $U(1)$
charges.
A very interesting related question is the construction of the string Bethe
equations describing states which in the undeformed case belong to
the same irreducible representation of $PSU(2,2|4)$.
The zero-mode states corresponding to fluctuations around the
$(J_1,J_2,J_3)_{\it vac}$ are only one of the simplest examples;
in general,  the Bethe solutions describing  extended multi-spin string
states in the same multiplet
as the highest-weight states  will exhibit similar  subtleties.
Their proper description appears to require a modification of the
standard thermodynamic limit arguments which should also be reflected in
the construction of the string Bethe equations.

\bigskip

There remain  many interesting  questions and directions for future
research.

It would be important to study this  non-supersymmetric
$\g_i$-deformed SYM  theory in detail, finding out,
in particular,   if it remains conformal (for properly adjusted
coupling and deformation parameters) even for finite $N$. In the
large $N$ limit this follows (to all orders in perturbation theory)
from the noncommutative nature of the deformation, the result of
\cite{filk} and the fact that $\g_i$ cannot be renormalized.\foot{
To see this we note that 
the operator \rf{starp} deforming the ordinary product must
have definite total dimension. Since the $U(1)$ generators have
vanishing anomalous dimensions, it follows that the same must hold for
$\g_i$.
The same argument implies that, in noncommutative field theories, the
$\theta$-parameter is  
not renormalized. Explicit calculations show that this is indeed
correct to two loops and general analysis of the renormalization of
such theories suggests that this is generally true.
}
At finite $N$ it is,  in principle,
straightforward to check the conformal invariance
(for properly adjusted parameters of the deformed Lagrangian)
to the first two  loop orders
using existing general relations for the  $\beta$-functions of
generic non-supersymmetric gauge theories
\ci{moc} (note that here, compared to orbifold models, all the fields
are in the
same adjoint representation of $U(N)$).
%(we expect it is).
The existence of exactly marginal non-supersymmetric  deformations
of $\N=4$ SYM theory implied  by  the AdS/CFT  duality
seems  an interesting subject worth detailed
study.\foot{Supersymmetric exactly
marginal deformations of  $\N=4$ SYM theory which can be obtained by
orbifolding were discussed  in \ci{raz}.}

\bigskip 
On the string theory side, it remains to construct the 
explicit form of the Green-Schwarz action describing the
$\g_i$-deformed theory. To do that  one may  
 apply the TsT transformation to the 
%Green-Schwarz $\k$-symmetric
superstring action on \adss \cite{MT} to 
%generate the $\g_i$-deformed
%background with all the supergravity fields included.
%find the explicit  form  of the Green-Schwarz  action 
%representing the $\g_i$-deformed theory
using the world-sheet  rules of T-duality formulated 
for the Green-Schwarz
superstring in \cite{RK}.\foot{Incidentally, that would  give the first
non-trivial example of the GS action in a non-supersymmetric background. 
To find the explicit form of this action  it  may be useful to start 
with the \adss action in a particular $\k$-symmetry 
gauge  where its fermionic structure is
explicit. One candidate for such a gauge is  
$(\Gamma_t + \Gamma_\phi)\theta=0$  where  $\phi$ 
is the direction which is T-dualized. 
}  
 %should  be useful for that).
%for solving this  problem. That would give the first nontrivial 
%example
%of the Green-Schwarz string on a nonsupersymmetric background.
The approach used in \cite{f} should then lead to
a local and periodic Lax representation for the complete
Green-Schwarz sigma model on the $\g_i$-deformed
background, related to the Lax pair for the \adss string \cite{BPR}
by the  TsT transformations.\foot{Again, a  more direct way to  derive the Lax pair for
 the GS string  may be to  start from the \adss  action 
  and do the transformations in the sigma
model rather than start with the sigma model  in the supergravity background
constructed using  the T-duality rules.}
Having found the Lax representation one  may then 
analyze the properties of the monodromy matrix and derive
the string Bethe equations for the $\g_i$-deformed model
analogous to those found  for superstring
on $AdS_5\times S^5$ in \cite{KMMZ,BKSZ}. The string Bethe
equations could then be
compared to  the thermodynamic limit of the Bethe equations
for the $\g_i$-deformed SYM theory
(this  was already  done 
%This program was successfully 
%carried out already 
%%It has been already done
for the simplest $su(2)_\g$ case in \cite{frt}).
One may also hope that the analysis of the string Bethe equations
will shed light on the structure of the dressing factor that appears
in the Bethe ansatz for quantum strings \cite{afs} (in the
 deformed case the dressing factor may
depend on $\gamma_i$ and that may lead to an additional 
 consistency condition for it).
 
\bigskip 

It would be  of much interest to study the   stability of this string theory, 
i.e. the presence of tachyons
in its spectrum. The  tachyons should be absent for
small enough $\g_i$  (as well as at the supersymmetric
point of equal $\g_i$).\foot{Since  the masses should be  smooth functions 
of the deformation  parameters $\g_i$, it  seems that 
  it is the lightest modes of the undeformed  background, 
i.e. the supergravity modes, that may become tachyonic first.
In general, there  may  be  a 
mixing between  ``momentum'' and ``winding'' modes 
under the TsT transformation (cf. the discussion of geodesics  in 
section 2).
There is some analogy with the case of the Melvin
twist of the flat-space theory \ci{rus}, where tachyons
appear in the winding sector for large enough  twist parameter 
$\g  > \g_{crit} = \ha w { R\ov \a'} $; tachyons   are present
for generic twist parameters, but are absent at  special
supersymmetric points. These winding tachyons can be seen at the supergravity
level  if one applies T-duality to the  flat Melvin background \ci{rus}.}

This deformed model may thus be useful   for understanding
 aspects  of closed-string tachyon physics in the AdS/CFT context
 (complementing orbifold model examples  like type 0 one  \ci{KT}
 with the advantage of having a tunable deformation parameter; an
interesting possibility is that
in the nonsupersymmetric $\gamma_i$-deformed theory
double-trace operators are not generated in perturbation
theory, cf. \cite{DKR}).
Some  particular  questions are  if tachyons are present in the
supergravity
 approximation  for generic 
 values of $\g_i$   and how to identify the corresponding operators
 on the gauge-theory side.

The present work gives also another illustration of
 the utility of the approach based on the
low-energy effective actions of Landau-Lifshitz type.
Another interesting problem (already mentioned in \ci{frt})
is to try to use the LL action found on  the gauge theory side
to reconstruct the dual geometry. This remains a challenge for
 the second
 $\N=1$ exactly marginal deformation of \ci{LEST}
 which preserves only one $U(1)$ isometry.\foot{For an attempt
  in this direction in the BMN limit  see \ci{NIPR};
  a perturbative supergravity approach  was developed in
  \ci{ahar}.}

%%
%SF
%%
   \bigskip

%%%%%%%%%%%%%%%%%%%%%%%%%%%%%%%%%%%%%%%%%
\section*{Acknowledgments }
%%%%%%%%%%%%%%%%%%%%%%%%%%%

We are  grateful to N. Beisert, M. Kruczenski,   
J. Michelson, 
 J. Russo  and K. Zarembo for  useful discussions and 
 important comments.
A.A.T.  acknowledges the hospitality of the Perimeter 
Institute
and the Fields Institute during the
completion  of this work. S.A.F. is
partially  supported by the EU-RTN network {\it Constituents, Fundamental
Forces and Symmetries of the Universe} (MRTN-CT-2004-005104).
The research of R.R. was
supported by the DOE grant No.~DE-FG02-91ER40671.
The work of A.A.T.  was supported  by the DOE grant
 DE-FG02-91ER40690
and also by the INTAS contract 03-51-6346 and 
the RS Wolfson award.

%%%%%%%%%%%%%%%%%%%%%%%%%%%%%%%%%%%%%%%%%%%%%%
\bigskip

\appendix

%  \section{%Appendix A:\

\def\lshcol{$\!\!\!\!\!\!\!$: }

\renewcommand{\thesection}{Appendix\ \Alph{section}}
\renewcommand{\theequation}{\Alph{section}.\arabic{equation}}
 \setcounter{equation}{0}

\section{\lshcol  Fluctuations near  ground states
 of spin chain
\label{app:massive}}

In section 2 we discussed
 fluctuations around vacua
  from the string theory perspective.
The fluctuations are found by expanding near 
the corresponding null geodesics representing
point-like string
states with lowest energy.
Here we shall attempt to analyze these
excitations from the gauge-theory
standpoint, using the one-loop Bethe ansatz.

There is a qualitative difference between the vacua
of the type
$(\JJ,0,0)$ and those of the type $(J_1,\,J_2,\,J_3)$:  the latter
are quantum states, corresponding to a nontrivial condensate of roots.
%Strangely enough, it turns out to be  harder to analyze the
%fluctuations around the first kind of vacua.

\bigskip

%%%%%%%%%%%%%%   %%%%%%%%%%%%%
\noindent
{\bf $\bullet$ Fluctuations near the $(\JJ,0,0)$ vacuum }
%%%%%%%%%%%%%%%%%%%%%%%%%%%%%%%%%%%%%%%%%%%
\bigskip

Let us first consider the small fluctuations
around the obvious
classical vacuum $(L,0,0)$ \ ($L \equiv \JJ$)
of the spin chain, i.e.
 consider the states with
 $J_2$ and $J_3$
being  small.\foot{The fluctuations around the other two
similar  vacua, $(0, L,0)$ and
$(0, 0, L)$, are related to those around the $(L,0,0)$
by simple relabeling.}

The rapidities of type $u_1$ in fact are divided into two groups. The
first group 
of rapidities which we denote 
$\uu12k\ (k=1,...,J_2)$
corresponds to fluctuations changing
the charge $J_2$, and the second
group of rapidities  
$\uu13k\ (k=1,...,J_3)$
corresponds to fluctuations changing
the charge $J_3$. The auxiliary rapidities 
$u_{2,k} \equiv \uu23k\ (k=1,...,J_3)$ 
are associated to rapidities 
$\uu13k$.

Following the intuition from the undeformed theory, we conclude that
if the number of excitations is small, their momenta are also small
and therefore their rapidities are large. To make this explicit
we introduce, as usual, the rescaled rapidities
\begin{eqnarray}
\xx12k=\frac{\uu12k}{L}~ , \ \ \ \   ~~~~~~~
\xx13k= \frac{\uu13k}{L}~~.
\label{scaling}
\end{eqnarray}
To get a consistent system of equations in the large $L$ limit we also have to assume that
the auxiliary rapidities $u_{2,k}$ have the following
 scaling behavior
\begin{eqnarray}
\uu23k = \uu13k + \sw23k = L \xx13k + \sw23k \ ,
\label{scaling2}
\end{eqnarray}
where 
$\sw23k$
do not depend on $L$. That means that in the large $L$ limit
an auxiliary rapidity 
$\uu23k$ 
may differ from
$\uu13k$ 
only by a constant.

Then, in terms of the rescaled rapidities the logarithm of the Bethe
equations 
(\ref{mainBetheSU3})-(\ref{auxBetheSU3}) and of the
momentum constraint (\ref{mom_constr}) expanded for large $L$ become
\begin{eqnarray}
-2\pi L\alpha_{21}+2\pi J_3(\alpha_{32}+\alpha_{21}+\alpha_{13})
-2\pi  n_{1,k}^{(2)}+
\frac{1}{\xx12k}&=&\sum_{\stackrel{i=1}{i\ne k}}^{J_2}
\frac{2/L}{\xx12k-\xx12i} + \sum_{i=1}^{J_3}
\frac{1/L}{\xx12k-\xx13i}
\ ,
\cr
-2\pi L\alpha_{21}+2\pi J_3(\alpha_{32}+\alpha_{21}+\alpha_{13})
+2\pi  n_{1,k}^{(3)}+
\frac{1}{\xx13k }&=&\sum_{\stackrel{i=1}{i\ne k}}^{J_3}
\frac{1/L}{\xx13k -\xx13i } + \sum_{i=1}^{J_2}
\frac{2/L}{\xx13k-\xx12i }\nonumber\\&-&i
\ln {\sw23k +i/2\over \sw23i -i/2}\ ,
\cr
2\pi L(\alpha_{21}+\alpha_{13}) -2\pi
(J_2+J_3)(\alpha_{32}+\a_{21}+\a_{13}) 
+2\pi  { n}_{2,k}^{(3)}
&=&\sum_{\stackrel{i=1}{i\ne k}}^{J_3}\frac{1/L}{\xx13k-\xx13i }
+
\sum_{i=1}^{J_2}\frac{1/L}{\xx12i -\xx13k } \nonumber\\
&+& i\ln {\sw23k +i/2 \over \sw23k -i/2}\ , \cr
2\pi L (J_2\alpha_{21}-J_3\alpha_{13})-2\pi m
&=&
\sum_{k=1}^{J_2}\frac{1}{\xx12k}
+\sum_{k=1}^{J_3}\frac{1}{\xx13k}\ ,
\label{longJJJbad}
\end{eqnarray}
where, as in \rf{alp},  $\a_{12} = - \g_3, \ \a_{23} = - \g_1, \
\a_{31} = - \g_2$ and $L\a_{ij}$ is assumed to be 
fixed in the scaling limit.
These equations hold regardless of any assumptions on the size of the
quantum numbers $J_i$. In the case of the $(J_1,J_2,J_3)$
fluctuations
around the
vacuum $(L,0,0)$ we further require that $J_1\sim {\cal O}(L)$ while
$J_2,J_3\sim {\cal O}(1)$. This assumption
implies that most of the terms in the equations
can be safely neglected, and the system takes the following simple form
\begin{eqnarray}
-2\pi L\alpha_{21}
-2\pi  n_{1,k}^{(2)}+
\frac{1}{\xx12k }&=&0\ ,
\cr
-2\pi L\alpha_{21}
+2\pi  n_{1,k}^{(3)}+
\frac{1}{\xx13k}&=&-i
\ln {\sw23k+i/2\over \sw23k-i/2}\ ,
\cr
2\pi L(\alpha_{21}+\alpha_{13})
+2\pi  n_{2,k}^{(3)}
&=&\hphantom{-}i \ln {\sw23k +i/2\over \sw23k -i/2}\ , \cr
2\pi L (J_2\alpha_{21}-J_3\alpha_{13})-2\pi m
&=&
\sum_{k=1}^{J_2}\frac{1}{\xx12k }+\sum_{k=1}^{J_3}\frac{1}{\xx13k}\ ,
\label{longJJJbad2}
\end{eqnarray}
where we  took into account that $\a_{ij}\sim 1/L$.

To find the energy spectrum we need to know only the rapidities 
$x_1^{(2)}$ and $x_1^{(3)}$ 
which can be easily determined from these equations
\begin{eqnarray}
\frac{1}{\xx12k }&=&\hphantom{-}2\pi \left( L\alpha_{21}+
 n_{1,k}^{(2)}\right) = 
\hphantom{-}2\pi \left(n_{1,k}^{(2)} +\g_3 L\right)
\ ,
\cr
\frac{1}{x_{1,k}^{(3)}}&=&-2\pi \left( L\alpha_{13}+
 n_{1,k}^{(3)}+{n}_{2,k}^{(3)}\right) 
= -2\pi \left( n_{1,k}^{(3)}+{n}_{2,k}^{(3)} +\g_2 L\right)
\ .
\label{longJJJbad3}
\end{eqnarray}
Shifting $n_1^{(3)}\to n_1^{(3)} - { n}_2^{(3)}$, 
we find the energy spectrum
\begin{eqnarray}
E=\frac{\lambda}{2L^2}\bigg[ \sum_{k=1}^{J_2}\left(n_{1,k}^{(2)}+
\g_3 L\right)^2 + \sum_{k=1}^{J_3}\left(n_{1,k}^{(3)}+
\g_2L\right)^2\bigg]\ ,
\label{enJ00}
\end{eqnarray}
which agrees precisely  with the leading term 
in the expansion of
the string theory result in \rf{fre}. Furthermore, the number of such states
is also correct, being equal to the number of states in the undeformed
theory.

The discussion above can be thought of as 
an explicit implementation of the general
arguments of \cite{STACK} regarding the structures appearing in the
thermodynamic limit of the ${\cal N}=4$  SYM spin chain.
 Adapting their
analysis to our context it follows that, for an 
arbitrary number of
excitations, the relevant equations in the thermodynamic limit 
are the first two equations (\ref{longJJJbad}) corresponding to 
the 1-stacks,
and the sum of  the second and third equations (\ref{longJJJbad}),
corresponding to the 2-stacks.\foot{The scaling \rf{scaling2}
implements the fact that the separation of roots inside a stack is
${\cal O}(1)$.}
Indeed, we have seen
that these combinations led to the solutions (\ref{longJJJbad3}).

\bigskip

%%%%%%%%%%%%%%%%%%%%%%%%%%%%%%%%%%%%%%%%%%
\noindent
{\bf $\bullet$ Fluctuations near the $(J_1,J_2,J_3)$  vacuum }
%%%%%%%%%%%%%%%%%%%%%%%%%%%%%%%%%%%%%%%%%%%
\bigskip

%\noindent
%$\bullet$

Let us now turn to the analysis of the fluctuations around
the quantum 
 vacuum $(J_1,J_2,J_3)$ \rf{parallel}, i.e. now 
  we will assume that $J_i\sim
{\cal O}(L)$ for all $i=1,2,3$. 
Since  such  states are built out of  large numbers of
excitations above the classical vacuum $(L,0,0)$ 
of the spin chain,  it is
convenient to take the thermodynamic limit,
i.e. $L \to \infty$. {}{}From the previous discussion and ref.\cite{STACK}
the relevant equations are a subset of (\ref{longJJJbad})
\begin{eqnarray}
%-2\pi L\alpha_{21}+2\pi J_3(\alpha_{32}+\alpha_{21}+\alpha_{13})
2\pi(J_3\alpha_{32}-J_1\alpha_{21}) +2\pi(J_3\a_{13}-J_2\a_{21})
-2\pi  n_{1,k}^{(2)}+
\frac{1}{\xx12k}&=&\sum_{\stackrel{i=1}{i\ne k}}^{J_2}
\frac{2/L}{\xx12k-\xx12i} + \sum_{i=1}^{J_3}
\frac{1/L}{\xx12k-\xx13i}
\ , \cr
%-2\pi L\alpha_{21}+2\pi J_3(\alpha_{32}+\alpha_{21}+\alpha_{13})
2\pi(J_3\alpha_{32}-J_1\alpha_{21}) +2\pi(J_3\a_{13}-J_2\a_{21})
+2\pi  n_{1,k}^{(3)}+
\frac{1}{\xx13k }&=&\sum_{\stackrel{i=1}{i\ne k}}^{J_3}
\frac{1/L}{\xx13k -\xx13i } + \sum_{i=1}^{J_2}
\frac{2/L}{\xx13k-\xx12i }\nonumber\\&-&
i\ln {\sw23k +i/2\over \sw23i -i/2}\ ,
\cr
%2\pi L \alpha_{13} -2\piJ_2(\alpha_{32}+\a_{21}+\a_{13}) 
2\pi(J_1\a_{13}-J_2\a_{32})+2\pi(J_3\a_{13}-J_2\a_{21})
+2\pi  {n}_{k}^{(3)}+\frac{1}{\xx13k}
&=&\sum_{\stackrel{i=1}{i\ne k}}^{J_3}\frac{2/L}{\xx13k-\xx13i }
+
\sum_{i=1}^{J_2}\frac{1/L}{\xx13k -\xx12i }  \cr
2\pi L (J_2\alpha_{21}-J_3\alpha_{13})-2\pi m L
\hphantom{~+\frac{1}{\xx13k}}
&=&
\sum_{k=1}^{J_2}\frac{1}{\xx12k}
+\sum_{k=1}^{J_3}\frac{1}{\xx13k}\ ,
\label{longJJJ}
\end{eqnarray}
with $n_k^{(3)}=n_{1,k}^{(3)}+n_{2,k}^{(3)}$.
As we have discussed in section 4, the vacuum $(J_1,J_2,J_3)_{ vac}$
exists whenever the angular momentum vector ${\ J}_{i, vac}$ is a zero
eigenvector of the deformation matrix $\alpha_{mn}$
(i.e. $\alpha_{12 } J_{2,vac} + \alpha_{13 } J_{3, vac}
= \alpha_{12 } J_{2,vac} - \alpha_{31 } J_{3, vac}=0$, etc.).
Fluctuations
around this vacuum have $j_i=J_i-J_{i,vac}\sim L^{\mu}$ with $\mu<1$ 
and therefore
the deformation-dependent terms on the left hand side of  the equations
above are of order $1/{L^{1-\mu}}$
(since $\a_{ij} L$ is fixed). 
 Such a source term appears in the equations
determining the vacuum rapidities as well and is an illustration of
the usual fact that excitations around any quantum vacuum back-react
on the vacuum condensate. In this case, the  deviation of the angular
momentum vector from being a 0-eigenvector of the deformation matrix
acts as a source in the equations for the vacuum rapidities and
renders them finite (albeit larger than the other ones by a factor of
$L$).

{}{}From the discussion in section 4 it is clear that
not all mode numbers are free parameters. Because of the fact that the
rapidities building the vacuum state  are infinite in the absence of 
additional excitations, the equations
(\ref{longJJJ}) imply that the corresponding mode numbers vanish.
We therefore have the following structure:
\begin{eqnarray}
\begin{array}{lllllll}
n_{1,k}& =0, &~k=1, \dots,  J_{2,vac}+J_{3,vac};  & &
 n_{2,k}&=0, &~k=1, \dots,  J_{3,vac}\cr
n_{1,k}&={\rm\scriptstyle free}, &~k=J_{2,vac}+J_{3,vac}+1, \dots,  J_{2}+J_{3};  & &
n_{2,k}&={\rm\scriptstyle free}, &~k=J_{3,vac}+1, \dots,  J_{3}
\end{array}
\end{eqnarray}
Further analyzing (\ref{longJJJ}) requires making a
distinction between the case in which all the mode numbers
 which are
free parameters are  nonzero and the case in which at least one of
the mode  numbers vanish. We will analyze here the first
case. 
%
% RR
%
and defer to \ref{app:zeromode}
the case of all vanishing mode numbers.

If all the free mode numbers are nonvanishing, it follows that 
the corresponding rapidities $x_{1,k}$ are of order unity as well as 
that in the corresponding equations the deformation-dependent terms 
are subleading compared to the mode numbers. Then, using the fact that 
all vacuum rapidities are large, it follows that
in the equations with nonvanishing mode number only very few
terms survive on the right hand side, insufficient to compensate for
the explicit $1/L$ suppression. This implies that $\xx12k$ and
$\xx13k$ are given by
\begin{eqnarray}
\frac{1}{\xx12k}={2\pi n_{1,k}^{(2)}}~,~~~~~\ \ \ \ \
\frac{1}{\xx13k}={2\pi n_{k}^{(3)}}\ , 
\end{eqnarray}
i.e.  are the same as in the undeformed theory.

The consistency of
all  other equations is also guaranteed by the fact that the
deformation enters at higher orders in the $1/L$ expansion.
If mode numbers of the auxiliary Bethe equations
are nonzero, they lead to quite complicated expressions for the
corresponding rapidities $u_2$. Fortunately,
 we do not need them since
they do not enter  the expressions for the 
energy or the momentum
constraint  given by:
\begin{eqnarray}
E=\frac{\lambda}{2L^2}\sum_{k=J_{2,vac}+J_{3,vac}+1}^{J_{2}+J_{3}}
n_{1,k}^2\ ,
~~~~~~~~~
\sum_{k=J_{2,vac}+J_{3,vac}+1}^{J_{2}+J_{3}}n_{1,k}=mL=0 \ . \la{hpu}
\end{eqnarray}
The vanishing of  the momentum number $m$ 
 is implied by the fact
that we considered only few excitations above the vacuum.

The conclusion is  that the small  fluctuations
around the $(J_1,J_2,J_3)$ vacuum having  non-zero mode numbers
are identical to those in the undeformed theory.
This is the same result as was 
 found on the string-theory side in
section 2.3.2.

\bigskip

 \setcounter{equation}{0}

\section{\lshcol  
The anomalous dimensions of operators dual to lowest 
energy pointlike strings
%Zero-mode fluctuations near the $(J_1,J_2,J_3)$  vacuum
%The gauge theory analysis of the zero modes
\label{app:zeromode}}

The special case in which all mode numbers in \rf{longJJJ} vanish is quite
interesting and nontrivial. The corresponding operators are BPS in the
absence of the deformation and thus their anomalous dimensions are
solely due to the presence of the deformation. In section 2.3.2 we
have seen that the zero-mode fluctuations around the
$(J_1,J_2,J_3)_{\it vac}$ geodesics are part of a larger class of
pointlike string configurations which in the large angular momentum
limit become (approximate) solutions. Their energies in this 
this limit are given by \rf{spi}. 
In this appendix we will go beyond the zero-mode approximation
$J_i/L\simeq J_{i,vac}/L$ and find the anomalous dimensions of 
the gauge theory operators corresponding to all such pointlike strings
captured by the deformed $su(3)$ sector.

There exists a conceptual issue related to 
the $(J_1,J_2,J_3)_{\it vac}$ state 
 being a {\it quantum rather than   classical  vacuum}. As mentioned
before, the Bethe equations \rf{mainBetheSU3}-\rf{auxBetheSU3} 
employ the state $(L,0,0)$ as the 
 vacuum 
 %ATT
 (``classical'' vacuum, i.e. a state with no excitations); 
  the quantum vacuum appears as a
nontrivial configuration of Bethe roots. {}From this perspective, 
fluctuations 
around this quantum state are on the one hand
 similar to a generic state with
large angular momenta and on the other hand special because they are
accidentally close to a zero energy state. 
While an analog of the Bethe equations
having  $(J_1,J_2,J_3)_{\it vac}$ as its ``classical'' 
vacuum would be  a desirable
starting point for studying the fluctuations near that state, 
deriving such  equations remains  an interesting open
problem.

In the following we will  discuss the excitations with 
vanishing mode numbers $n_k$ close to the $(J_1,J_2,J_3)_{\it vac}$
state, using the Bethe equations \rf{mainBetheSU3}-\rf{auxBetheSU3}. 
%Perhaps surprisingly, 
Remarkably, we  will find that the results agree with the exact
string theory predictions (\ref{spi}). 

Depending on the departure from the  $(J_1,J_2,J_3)_{\it vac}$  
state, it is easy
to see what is the scaling of the Bethe roots with the length of the
chain. The relevant equations follow from  \rf{longJJJ}
\begin{eqnarray}
%-2\pi L\alpha_{21}+2\pi J_3(\alpha_{32}+\alpha_{21}+\alpha_{13})
%%2\pi(J_3\alpha_{32}-J_1\alpha_{21}) +2\pi(J_3\a_{13}-J_2\a_{21})
2\pi j_3 \gga +
\frac{1}{\xx12k}&=&\sum_{\stackrel{i=1}{i\ne k}}^{J_2}
\frac{2/L}{\xx12k-\xx12i} + \sum_{i=1}^{J_3}
\frac{1/L}{\xx12k-\xx13i}
\ , 
\label{z1}\\
%2\pi L \alpha_{13} -2\piJ_2(\alpha_{32}+\a_{21}+\a_{13}) 
%2\pi(J_1\a_{13}-J_2\a_{32})+2\pi(J_3\a_{13}-J_2\a_{21})
-2\pi j_2\gga
+\frac{1}{\xx13k}
&=&\sum_{\stackrel{i=1}{i\ne k}}^{J_3}\frac{2/L}{\xx13k-\xx13i }
+
\sum_{i=1}^{J_2}\frac{1/L}{\xx13k -\xx12i }  \ , 
\label{z2}\\
2\pi (j_2\g_3-j_3\g_2)
\hphantom{~+\frac{1}{\xx13k}}
&=&
\sum_{k=1}^{J_2}\frac{1/L}{\xx12k}
+\sum_{k=1}^{J_3}\frac{1/L}{\xx13k}\ , 
\label{longJJJ_zero}
\end{eqnarray}
where we explicitly used the expression of the vacuum quantum numbers
$(J_1,J_2,J_3)_{\it vac}$ and, as before,
\begin{equation}\la{je}
j_i=J_i-\frac{\g_i}{\gga}L~,  \ \ \ \ \ ~~~~~~\gga=\g_1+\g_2+\g_3~~.
\end{equation}
The parameters $j_i$ describe the deviation of the state
$(J_1,J_2,J_3)$ from the vacuum; therefore, their scaling 
with the length of the chain is $j\sim L^\mu$ with 
$0\le\mu\le 1$. Then, from \rf{longJJJ_zero} it trivially follows that
%
%, if the deviation
%from vacuum scales like $j\sim L^\mu$, then 
%
the constant source terms
scale like $j\gga\sim  L^{\mu-1}$ which leads to rapidities
$\xx12k,\,\xx13k\sim L^{1-\mu}$.
Still, to leading order, the expression for  the energy does not involve
any  fractional powers of $L$.
% in spite of the this dependence. 
Indeed, the energy is an even function
%power 
of the
constant source in the Bethe equations, which vanishes in its
absence. Thus, schematically and  to the leading order, the energy 
behaves as
\begin{eqnarray}
E&=&\frac{\lambda}{8\pi^2}\sum_{k=1}^{J_2+J_3}
\frac{1}{u_{1,k}^2+\textstyle{\frac{1}{4}}}
=\frac{\lambda}{8\pi^2L^{2(2-\mu)}}F({x}_{1}^{(2)}, {x}_{1}^{(3)})\cr
& \sim & 
\frac{\lambda}{8\pi^2L^{2(2-\mu)}}(\gamma j L^{1-\mu})^2
\left[{\cal O}(J_2+J_3)+\dots\right]
\sim \frac{\tilde \lambda}{8\pi^2}(\gamma L)^2 
\frac{j^2}{L}
\left[{\cal O}(1)+\dots\right] \ . 
\label{trace_mu}
\end{eqnarray}
This $L$-dependence is similar to the one derived on the 
string theory side in \rf{spe}. The existence of a rescaling of the
rapidities which makes all terms in the Bethe equations of the same
order also implies that we can safely neglect terms of the type
$1/(Lx)$. This observation will be useful shortly.

The equations (\ref{z1})-(\ref{longJJJ_zero}) are similar to 
those in \cite{FrKr}. The differences are the
nonintegrality of the constant term on their left hand side and 
potential term on the left hand side of (\ref{z2}). 
The solution is,  however,  similar to that of \cite{FrKr}. 
To analyze them it is useful to proceed 
in the standard way and introduce the
resolvents
\begin{eqnarray}
G_{i}(x)=\frac{1}{L}\sum_{k=1}^{J_{i}}\frac{1}{x-\xx1ik}~,
 \ \ \ \ ~~~~~~i=2,3~, 
\label{resolv}
\end{eqnarray}
in terms of which the anomalous dimensions are
\begin{eqnarray}
E=-\frac{\l}{8\pi^2 L}\,(G'_2(0)+G'_3(0))~~.
\end{eqnarray}

To find $G'_2(0)$ and $G'_3(0)$ we begin by
multiplying the first equation in (\ref{z1})-(\ref{longJJJ_zero})
by $\frac{1}{x-\xx12k}$ and the second
by $\frac{1}{x-\xx13k}$ and summing all the equations. This leads to
\begin{eqnarray}
2\pi j_3\gga G_2(x) + \frac{1}{x}\left(G_2(x)-G_2(0)\right)&=&G_2(x)^2
-\frac{1}{L}G'_2(x)+
\frac{1}{L^2}\sum_{k=1}^{J_2}\sum_{i=1}^{J_3}
\frac{1}{(x-\xx12k)(\xx12k-\xx13i)}\cr
-2\pi j_2\gga G_3(x) + \frac{1}{x}\left(G_3(x)-G_3(0)\right)&=&G_3(x)^2
-\frac{1}{L}G'_3(x)+
\frac{1}{L^2}\sum_{k=1}^{J_3}\sum_{i=1}^{J_2}
\frac{1}{(x-\xx13k)(\xx13k-\xx12i)}\cr
-2\pi(j_2\g_3-j_3\g_2)&=&G_2(0)+G_3(0)~.
\end{eqnarray}
Further summing the first 
 two equations and neglecting subleading terms
we find
\begin{eqnarray}
(2\pi j_3\gga) G_2(x) +(-2\pi j_2\gga) G_3(x)&+&
\frac{1}{x} \left[G_2(x)+G_3(x)-G_2(0)-G_3(0)\right] \ , \cr
&=&[G_2(x)+G_3(x)]^2-G_2(x)G_3(x)
%%-\frac{1}{L}G'(x)
\cr
-2\pi(j_2\g_3-j_3\g_2)&=&G_2(0)+G_3(0)~~.
\end{eqnarray}
The limit $x\rightarrow 0$ expresses the derivative of the sum of
the resolvents evaluated at the origin 
in terms of the values of $G_2$ and $G_3$ at $x=0$:
\begin{eqnarray}
\label{dG}
G'_2(0)+G'_3(0)&=&~~C^2-G_2(0)G_3(0)-A G_2(0) -BG_3(0)\ , \\
G_2(0)+G_3(0)&=&-C \ , 
\label{mom_const_B}
\end{eqnarray}
where, to shorten later equations, we introduced the notation
\begin{eqnarray}
A=2\pi j_3\gga~ ,\ \ \ \  ~~~~~~B=-2\pi j_2\gga
~, \ \ \ \ \ ~~~~
C=2\pi(j_2\g_3-j_3\g_2)\ . 
\label{def}
\end{eqnarray}
%Two  
To find $G_2(0)$ and $G_3(0)$ we need another equation 
in addition to \rf{mom_const_B}; it can 
be obtained by first
multiplying (\ref{z1}) by the factor 
$\sum_{m=1}^{J_3}\frac{1/L^2}{\xx13m-\xx12k}$
%%$1/L^2 \sum_{m=1}^{J_3}1/(\xx13m-\xx12k)$
and summing over $k$,  and by multiplying (\ref{z2}) by 
$\sum_{m=1}^{J_2}\frac{1/L^2}{\xx12m-\xx13k}$
%%$1/L^2 \sum_{m=1}^{J_2}1/(\xx12m-\xx13k)$ 
and summing over $k$:
\begin{eqnarray}
&&A\sum_{m=1}^{J_3}\sum_{k=1}^{J_2}\frac{1}{\xx13m-\xx12k}+
\sum_{m=1}^{J_3}\sum_{k=1}^{J_2}\frac{1}{\xx12k(\xx13m-\xx12k)}
\cr
&&
=
\sum_{m=1}^{J_3}\sum_{k\ne i=1}^{J_2}
\frac{1/L}{(\xx13m-\xx12k)(\xx13m-\xx12i)}+
\sum_{m,i=1}^{J_3}\sum_{k=1}^{J_2}
\frac{1/L}{(\xx13m-\xx12k)(\xx12k-\xx13i)}
\cr
&&B\sum_{m=1}^{J_2}\sum_{k=1}^{J_3}\frac{1}{\xx12m-\xx13k}-
\sum_{m=1}^{J_2}\sum_{k=1}^{J_3}\frac{1}{\xx13k(\xx13k-\xx12m)}\cr
&&
=
-\sum_{m=1}^{J_2}\sum_{k\ne i=1}^{J_3}
\frac{1/L}{(\xx13k-\xx12m)(\xx12m-\xx13i)}-
\sum_{m,i=1}^{J_2}\sum_{k=1}^{J_3}
\frac{1/L}{(\xx13k-\xx12m)(\xx13k-\xx12i)}  
\end{eqnarray}
Then, summing these two equations and dividing by $L^2$ leads to 
\begin{eqnarray}
(A-B)
\sum_{m=1}^{J_3}\sum_{k=1}^{J_2}\frac{1/L^2}{\xx13m-\xx12k}
+G_2(0)G_3(0)=0\ . 
\end{eqnarray}
The unknown sum can be determined by summing the equations (\ref{z1})
or the equations (\ref{z2}):
\begin{eqnarray}
\sum_{m=1}^{J_3}\sum_{k=1}^{J_2}\frac{1/L^2}{\xx13m-\xx12k}&=&
-G_3(0)+B\bfa_2=G_2(0)-A\bfa_3\cr
&=&-G_3(0)+B\left(\frac{\g_2}{\gga}+\frac{j_2}{L}\right)
=G_2(0)-A\left(\frac{\g_3}{\gga}+\frac{j_3}{L}\right)
\end{eqnarray}
where we used the notation 
$$\bfa_i\equiv {J_i\ov L}={\g_i\ov\gga}+
{j_i\ov L} $$
  for  the filling fractions.
Thus, the $G_2(0)$ and $G_3(0)$ are determined by:
\begin{eqnarray}
(A-B)[B\bfa_3-G_3(0)]+G_2(0)G_3(0)&=&0\ , \cr
G_2(0)+G_3(0)&=&-C\ . 
\end{eqnarray}
{}From the definition \rf{resolv} it follows that resolvents $G_2$ and $G_3$
identically vanish if $J_2=J_3=0$; we will therefore
pick the solution for $G_2(0)$ and $G_3(0)$ which also 
vanishes in this limit:
\begin{eqnarray}
G_2(0)&=&\frac{1}{2}\left[~~A-B-C - 
\sqrt{4\, \bfa_3\, B(A-B)+(A-B+C)^2} \right]\ , \cr
G_3(0)&=&\frac{1}{2}\left[-A+B-C +
\sqrt{4\, \bfa_3\, B(A-B)+(A-B+C)^2} \right]
\end{eqnarray}
Using  (\ref{dG}), it is then easy to find what $G_2'(0)+G_3'(0)$ is:
\begin{eqnarray}
G_2'(0)+G_3'(0)=
%%-\a_3 B^2 + C^2 + A(\a_3 B + C)~~.
\bfa_3 \, B(A-B)+C(A+C)\ . 
\label{dGsol}
\end{eqnarray}
Finally, using the definitions (\ref{def}) to express (\ref{dGsol}) in
terms of the deformation parameters $\g_i$ and 
the deviations of the angular momenta from the vacuum values 
 $j_i$,  
the anomalous dimensions are found to be 
\begin{eqnarray}
E=-\frac{\l}{8\pi^2 L}(G'_2(0)+G'_3(0))=
\frac{\l}{2
L}\left[\g_1\g_2j_3^2+\g_2\g_3j_1^2+\g_3\g_1j_2^2
-\frac{1}{L}(\g_1+\g_2+\g_3)^2j_1j_2j_3\right] 
\label{Bspe}
\end{eqnarray}
As promised, this reproduces the exact 
string theory result (\ref{spi}).

The calculation 
above shows that there exists a configuration of Bethe roots whose
energy matches that of the string theory zero modes. 
Even though we have not found
explicitly the rapidity distribution (since we only needed the values
of the resolvents and their first derivative at the origin) we may
comment on some of its features. In the undeformed theory the
$(J_1,J_2,J_3)$ 
BPS states are described by infinite rapidities which are also
infinitely separated. As we turn on the deformation, the Bethe roots
$u_{1,2}$ 
descend to  finite distance, of the order of $L/(j\g)$. 
The fact that
initially their differences were also infinite suggests that in the
presence of the deformation they will also be of the order of
$L/(j\g)$. The 
distance between them is still  large, and  that suggests that 
the Bethe roots describing the zero modes do not condense. 

Besides the solution described above, the equations
\rf{z1}-\rf{longJJJ_zero} have additional ones. For example, 
if $j_2=j_3$ and $\g_2=\g_3$ it is possible to construct a solution
satisfying $G_2(x)-G_3(x)={\cal C} x (G_2(x)+G_3(x))$ where ${\cal C}$ 
is a constant which may be determined from the asymptotic behavior of
the resolvents. It turns out that $G_2(x)+G_3(x)$ has no cut, so it
also does not describe a root condensate. Rather, it has two poles, at
$\pm i (\sqrt{3} {\cal C})^{-1}$. Its energy has the same scaling with
the length of the chain as in \rf{trace_mu}, but it is a nonanalytic
function of the deformation parameters $\g_i$.  This feature might
tempt one to discard it, based on the fact that
 the undeformed theory
should be reached smoothly in 
the limit  $\g_i \to 0$. The physical interpretation 
of this solution is  not clear at the moment. 

It is worth pointing out that, in the calculation and the matching 
described above, the value of the power $0\le\mu\le 1$ 
in the scaling $j\sim L^\mu$ was unimportant. This is in agreement
with the string theory discusion in  section 2.3.2. {}{}From the perspective of
the Bethe ansatz we expect finite size corrections to the energies
\rf{Bspe}. Since the equations \rf{z1}-\rf{longJJJ_zero} include all
terms up to $O(1)$ in the $1/L$ expansion of 
the logarithm of the
Bethe equations,  these corrections should be suppressed by
additional powers of $1/L$. 
It would be interesting, though appears to be quite challenging, 
to compare
these corrections to the $\a'$ corrections
 on the string theory side.
Techniques developed in \ci{BEFR} may be
 useful in this respect.

%\newpage

%%%%%%%%%%%%%%%%%%%%%%%%%%%%%%%

\begin{thebibliography}{99}



\bi{LM}
  O.~Lunin and J.~Maldacena,
  ``Deforming field theories with U(1) x U(1) global symmetry and
    their gravity duals,''
  hep-th/0502086.
  %%CITATION = HEP-TH 0502086;%%

\bi{frt}
S.~A.~Frolov, R.~Roiban and A.~A.~Tseytlin,
  ``Gauge - string duality for superconformal deformations of N = 4 super
  Yang-Mills theory,''
  hep-th/0503192.
  %%CITATION = HEP-TH 0503192;%%

\bi{f}
S.~Frolov,
  ``Lax pair for strings in Lunin-Maldacena background,''
  JHEP {\bf 0505}, 069 (2005)
  [hep-th/0503201].
  %%CITATION = HEP-TH 0503201;%%

\bi{kas}
 S.~Kachru and E.~Silverstein,
  ``4d conformal theories and strings on orbifolds,''
  Phys.\ Rev.\ Lett.\  {\bf 80}, 4855 (1998)
  [hep-th/9802183].
  %%CITATION = HEP-TH 9802183;%%
A.~E.~Lawrence, N.~Nekrasov and C.~Vafa,
  ``On conformal field theories in four dimensions,''
  Nucl.\ Phys.\ B {\bf 533}, 199 (1998)
  [hep-th/9803015].
  %%CITATION = HEP-TH 9803015;%%
 M.~Bershadsky, Z.~Kakushadze and C.~Vafa,
  ``String expansion as large N expansion of gauge theories,''
  Nucl.\ Phys.\ B {\bf 523}, 59 (1998)
  [hep-th/9803076].
  %%CITATION = HEP-TH 9803076;%%


\bi{KT}
I.~R.~Klebanov and A.~A.~Tseytlin,
  ``A non-supersymmetric large N CFT from type 0 string theory,''
  JHEP {\bf 9903}, 015 (1999)
  [hep-th/9901101].
  %%CITATION = HEP-TH 9901101;%%
I.~R.~Klebanov,
  ``Tachyon stabilization in the AdS/CFT correspondence,''
  Phys.\ Lett.\ B {\bf 466}, 166 (1999)
  [hep-th/9906220].
  %%CITATION = HEP-TH 9906220;%%





\bi{cot}
F. Bigazzi, A.L.~Cotrone, L.~Girardello and A.~Zaffaroni,
  ``pp- wave and non supersymmetric gauge theory,''
  JHEP {\bf 0210}, 030 (2002)
  [hep-th/0205296].
  %%CITATION = HEP-TH 0205296;%%


\bi{ft2}
S.~Frolov and A.~A.~Tseytlin, ``Multi-spin string solutions in
\adss,'' Nucl.\ Phys.\ B {\bf 668}, 77 (2003) [hep-th/0304255].
%%CITATION = HEP-TH 0304255;%%
``Semiclassical quantization of rotating superstring in AdS(5) x S(5),''
  JHEP {\bf 0206}, 007 (2002)
  [hep-th/0204226].
  %%CITATION = HEP-TH 0204226;%%

\bi{minah}
J.~A.~Minahan,
  ``Circular semiclassical string solutions on AdS(5) x S(5),''
  Nucl.\ Phys.\ B {\bf 648}, 203 (2003)
  [hep-th/0209047].
  %%CITATION = HEP-TH 0209047;%%

\bi{art}
G.~Arutyunov, J.~Russo and A.~A.~Tseytlin, ``Spinning strings in
\adss: New integrable system relations,'' Phys.\ Rev.\ D {\bf 69},
086009 (2004) [hep-th/0311004].
%%CITATION = HEP-TH 0311004;%%
G.~Arutyunov, S.~Frolov, J.~Russo and A.~A.~Tseytlin,
``Spinning strings in AdS(5) x S5 and integrable systems,''
Nucl.\ Phys.\ B {\bf 671}, 3 (2003)
[hep-th/0307191].
%%CITATION = HEP-TH 0307191;%%


\bi{BMN}
D.~Berenstein, J.~M.~Maldacena and H.~Nastase, ``Strings
in flat space and pp waves {}from N =4 super Yang Mills,'' JHEP
{\bf 0204}, 013 (2002) [hep-th/0202021].
%%CITATION = HEP-TH 0202021;%%

\bibitem{NIPR}
  V.~Niarchos and N.~Prezas,
  ``BMN operators for N = 1 superconformal Yang-Mills theories and
    associated string backgrounds,''
  JHEP {\bf 0306}, 015 (2003)
  [hep-th/0212111].
  %%CITATION = HEP-TH 0212111;%%


\bi{mat}
T.~Mateos,
  ``Marginal deformation of N = 4 SYM and Penrose limits with continuum
  spectrum,''
  hep-th/0505243.
  %%CITATION = HEP-TH 0505243;%%

\bi{koch}
R.~de Mello Koch, J.~Murugan, J.~Smolic and M.~Smolic,
  ``Deformed PP-waves from the Lunin-Maldacena background,''
  hep-th/0505227.
  %%CITATION = HEP-TH 0505227;%%

\bi{KRUC}
  M.~Kruczenski,
  ``Spin chains and string theory,''
  Phys.\ Rev.\ Lett.\  {\bf 93}, 161602 (2004)
  [hep-th/0311203].
  %%CITATION = HEP-TH 0311203;%%
  M.~Kruczenski, A.~V.~Ryzhov and A.~A.~Tseytlin,
  ``Large spin limit of AdS(5) x S5 string theory and low energy
expansion of ferromagnetic spin chains,''
  Nucl.\ Phys.\ B {\bf 692}, 3 (2004)
  [hep-th/0403120].
  %%CITATION = HEP-TH 0403120;%%

\bi{kt}
M.~Kruczenski and A.~A.~Tseytlin,
  ``Semiclassical relativistic strings in S**5 and long coherent operators in N
  = 4 SYM theory,''
  JHEP {\bf 0409}, 038 (2004)
  [hep-th/0406189].
  %%CITATION = HEP-TH 0406189;%%

\bi{hlo}
R.~Hernandez and E.~Lopez,
``The SU(3) spin chain sigma model and string theory,'' JHEP {\bf
0404}, 052 (2004) [hep-th/0403139].
%%CITATION = HEP-TH 0403139;%%

\bi{st}B.~J.~Stefanski and A.~A.~Tseytlin, ``Large spin limits of
AdS/CFT and generalized Landau-Lifshitz equations,'' JHEP {\bf 0405},
042 (2004) [hep-th/0404133].
%%CITATION = HEP-TH 0404133;%%


\bibitem{RR}
  R.~Roiban,
  ``On spin chains and field theories,''
  JHEP {\bf 0409}, 023 (2004)
  [hep-th/0312218].
  %%CITATION = HEP-TH 0312218;%%

\bibitem{BECH}
  D.~Berenstein and S.~A.~Cherkis,
  ``Deformations of N = 4 SYM and integrable spin chain models,''
  Nucl.\ Phys.\ B {\bf 702}, 49 (2004)
  [hep-th/0405215].
  %%CITATION = HEP-TH 0405215;%%


\bibitem{LEST}
  R.~G.~Leigh and M.~J.~Strassler,
  ``Exactly marginal operators and duality in four-dimensional N=1
  supersymmetric gauge theory,''
  Nucl.\ Phys.\ B {\bf 447}, 95 (1995)
  [hep-th/9503121].
  %%CITATION = HEP-TH 9503121;%%

\bi{old}
A.~Parkes and P.~C.~West,
  ``Finiteness In Rigid Supersymmetric Theories,''
  Phys.\ Lett.\ B {\bf 138}, 99 (1984).
  %%CITATION = PHLTA,B138,99;%%
  ``Three Loop Results In Two Loop Finite Supersymmetric Gauge Theories,''
  Nucl.\ Phys.\ B {\bf 256}, 340 (1985).
  %%CITATION = NUPHA,B256,340;%%
D.R.T.~Jones and L.~Mezincescu,
  ``The Chiral Anomaly And A Class Of Two Loop Finite Supersymmetric Gauge
  Theories,''
  Phys.\ Lett.\ B {\bf 138}, 293 (1984).
  %%CITATION = PHLTA,B138,293;%%
D.R.T.~Jones and A.~J.~Parkes,
  ``Search For A Three Loop Finite Chiral Theory,''
  Phys.\ Lett.\ B {\bf 160}, 267 (1985).
  %%CITATION = PHLTA,B160,267;%%



\bi{mz}
J.~A.~Minahan and K.~Zarembo,
 ``The Bethe-ansatz for N = 4 super Yang-Mills,''
 JHEP {\bf 0303}, 013 (2003)
 [hep-th/0212208].
 %%CITATION = HEP-TH 0212208;%%


\bibitem{BDS}
N.~Beisert, V.~Dippel and M.~Staudacher, ``A novel long range spin
chain and planar N = 4 super Yang-Mills,'' JHEP {\bf 0407}, 075
(2004) [hep-th/0405001].
%%CITATION = HEP-TH 0405001;%%

\bi{Mina}
J.~Engquist, J.~A.~Minahan and K.~Zarembo,
 ``Yang-Mills duals for semiclassical strings on AdS(5) x S5,''
 JHEP {\bf 0311}, 063 (2003)
 [hep-th/0310188].
 %%CITATION = HEP-TH 0310188;%%


\bibitem{GKP2}
S.~S.~Gubser, I.~R.~Klebanov and A.~M.~Polyakov,  ``A
semi-classical limit of the gauge/string correspondence,'' Nucl.\
Phys.\ B {\bf 636} (2002) 99,  hep-th/0204051.


\bi{pope}
M.~Cvetic, H.~Lu, C.~N.~Pope and K.~S.~Stelle,
  ``Linearly-realised worldsheet supersymmetry in pp-wave background,''
  Nucl.\ Phys.\ B {\bf 662}, 89 (2003)
  [hep-th/0209193].
  %%CITATION = HEP-TH 0209193;%%

\bi{blau}
M.~Blau, M.~O'Loughlin, G.~Papadopoulos and A.~A.~Tseytlin,
  ``Solvable models of strings in homogeneous plane wave backgrounds,''
  Nucl.\ Phys.\ B {\bf 673}, 57 (2003)
  [hep-th/0304198].
  %%CITATION = HEP-TH 0304198;%%


\bi{mets}
 R.~R.~Metsaev and A.~A.~Tseytlin,
  ``Exactly solvable model of superstring in plane wave Ramond-Ramond
  background,''
  Phys.\ Rev.\ D {\bf 65}, 126004 (2002)
  [hep-th/0202109].
  %%CITATION = HEP-TH 0202109;%%
 J.~G.~Russo and A.~A.~Tseytlin,
  ``On solvable models of type IIB superstring in NS-NS and R-R plane wave
  backgrounds,''
  JHEP {\bf 0204}, 021 (2002)
  [hep-th/0202179].
  %%CITATION = HEP-TH 0202179;%%



\bi{ft3}
S.~Frolov and A.~A.~Tseytlin,
``Quantizing three-spin string solution in \adss,'' JHEP {\bf
0307}, 016 (2003) [hep-th/0306130].
%%CITATION = HEP-TH 0306130;%%



\bi{quant}
M.~P.~Bellon and M.~Talon,
  ``Spectrum of the quantum Neumann model,''
  Phys.\ Lett.\ A {\bf 337}, 360 (2005)
  [hep-th/0407005].
  %%CITATION = HEP-TH 0407005;%%
  ``Separation of variables for the classical and quantum Neumann model,''
  Nucl.\ Phys.\ B {\bf 379}, 321 (1992)
  [hep-th/9201035].
  %%CITATION = HEP-TH 9201035;%%
A.~J.~Macfarlane,
  ``The Quantum Neumann model with the potential of Rosochatius,''
  Nucl.\ Phys.\ B {\bf 386}, 453 (1992).
  %%CITATION = NUPHA,B386,453;%%

\bi{ide}
K.~Ideguchi,
``Semiclassical strings on AdS(5) x S5/Z(M) and operators in
orbifold field theories,''
JHEP {\bf 0409}, 008 (2004) [hep-th/0408014].
%%CITATION = HEP-TH 0408014;%%

\bibitem{BR}
  N.~Beisert and R.~Roiban,
  ``Beauty and the twist: The Bethe ansatz for twisted N = 4 SYM,''
  hep-th/0505187.
  %%CITATION = HEP-TH 0505187;%%

\bibitem{BELE}
  D.~Berenstein, V.~Jejjala and R.~G.~Leigh,
  ``Marginal and relevant deformations of N = 4 field theories and
  non-commutative moduli spaces of vacua,''
  Nucl.\ Phys.\ B {\bf 589}, 196 (2000)
  [hep-th/0005087].
  %%CITATION = HEP-TH 0005087;%%
  D.~Berenstein and R.~G.~Leigh,
``Discrete torsion, AdS/CFT and duality,''
  JHEP {\bf 0001}, 038 (2000)
  [hep-th/0001055].
  %%CITATION = HEP-TH 0001055;%%

\bi{bob}
N.~P.~Bobev, H.~Dimov and R.~C.~Rashkov,
  ``Semiclassical strings in Lunin-Maldacena background,''
  hep-th/0506063.
  %%CITATION = HEP-TH 0506063;%%

\bibitem{dor}
  N.~Dorey and T.~J.~Hollowood,
  ``On the Coulomb branch of a marginal deformation of N = 4 SUSY Yang-Mills,''
  hep-th/0411163.
  %%CITATION = HEP-TH 0411163;%%


\bi{filk}
T.~Filk,
  ``Divergencies in a field theory on quantum space,''
  Phys.\ Lett.\ B {\bf 376}, 53 (1996).
  %%CITATION = PHLTA,B376,53;%%

\bi{rum}
J.~M.~Maldacena and J.~G.~Russo,
  ``Large N limit of non-commutative gauge theories,''
  JHEP {\bf 9909}, 025 (1999)
  [hep-th/9908134].
  %%CITATION = HEP-TH 9908134;%%



\bi{guf}
D.Z. Freedman and U.  G\"ursoy,
``Comments on $\b$-deformed $\N=4$ SYM theory'',
hep-th/0506128.

\bi{zan}
S.~Penati, A.~Santambrogio and D.~Zanon,
  ``Two-point correlators in the beta-deformed N=4 SYM at the next-to-leading
  order,''
  hep-th/0506150.
  %%CITATION = HEP-TH 0506150;%%

\bi{bmsz}
N.~Beisert, J.~A.~Minahan, M.~Staudacher and K.~Zarembo,
  ``Stringing spins and spinning strings,''
  JHEP {\bf 0309}, 010 (2003)
  [hep-th/0306139].
  %%CITATION = HEP-TH 0306139;%%



\bi{raz}
 O.~Aharony and S.~S.~Razamat,
  ``Exactly marginal deformations of N = 4 SYM and of its supersymmetric
  orbifold descendants,''
  JHEP {\bf 0205}, 029 (2002)
  [hep-th/0204045].
  %%CITATION = HEP-TH 0204045;%%
S.~S.~Razamat,
  ``Marginal deformations of N = 4 SYM and of its supersymmetric orbifold
  descendants,''
  hep-th/0204043.
  %%CITATION = HEP-TH 0204043;%%

\bi{rus}
J.~G.~Russo and A.~A.~Tseytlin,
  ``Magnetic flux tube models in superstring theory,''
  Nucl.\ Phys.\ B {\bf 461}, 131 (1996)
  [hep-th/9508068].
  ``Supersymmetric fluxbrane intersections and closed string tachyons,''
  JHEP {\bf 0111}, 065 (2001)
  [hep-th/0110107].
  %%CITATION = HEP-TH 0110107;%%
  ``Magnetic backgrounds and tachyonic instabilities in closed superstring
  theory and M-theory,''
  Nucl.\ Phys.\ B {\bf 611}, 93 (2001)
  [hep-th/0104238].
  %%CITATION = HEP-TH 0104238;%%
  
  
\bi{moc}
I.~Jack and H.~Osborn,
``Two Loop Background Field Calculations For Arbitrary Background Fields,''
  Nucl.\ Phys.\ B {\bf 207}, 474 (1982).
  %%CITATION = NUPHA,B207,474;%%
  ``General Two Loop Beta Functions For Gauge Theories With Arbitrary Scalar
  %Fields,''
  J.\ Phys.\ A {\bf 16}, 1101 (1983).
  %%CITATION = JPAGB,A16,1101;%%
 ``General Background Field Calculations With Fermion Fields,''
  Nucl.\ Phys.\ B {\bf 249}, 472 (1985).
  %%CITATION = NUPHA,B249,472;%%  
M.~E.~Machacek and M.~T.~Vaughn,
``Two Loop Renormalization Group Equations In A General Quantum Field Theory.
  1. Wave Function Renormalization,''
  Nucl.\ Phys.\ B {\bf 222}, 83 (1983).
  %%CITATION = NUPHA,B222,83;%%
``Two Loop Renormalization Group Equations In A General Quantum Field Theory.
  2. Yukawa Couplings,''
  Nucl.\ Phys.\ B {\bf 236}, 221 (1984).
  %%CITATION = NUPHA,B236,221;%%
 ``Two Loop Renormalization Group Equations In A General Quantum Field
Theory. 3. Scalar Quartic Couplings,''
  Nucl.\ Phys.\ B {\bf 249}, 70 (1985).
  %%CITATION = NUPHA,B249,70;%%

\bi{kw}
I.~R.~Klebanov and E.~Witten,
  ``Superconformal field theory on threebranes at a Calabi-Yau
singularity,''
  Nucl.\ Phys.\ B {\bf 536}, 199 (1998)
  [hep-th/9807080].
  %%CITATION = HEP-TH 9807080;%%

\bi{DKR}
  A.~Dymarsky, I.~R.~Klebanov and R.~Roiban,
  ``Perturbative search for fixed lines in large N gauge theories,''
  hep-th/0505099.
  %%CITATION = HEP-TH 0505099;%%

\bi{ahar}
 O.~Aharony, B.~Kol and S.~Yankielowicz,
  ``On exactly marginal deformations of N = 4 SYM and type IIB  supergravity on
  AdS(5) x S5,''
  JHEP {\bf 0206}, 039 (2002)
  [hep-th/0205090].
  %%CITATION = HEP-TH 0205090;%%

\bibitem{MT}
  R.~R.~Metsaev and A.~A.~Tseytlin,
  ``Type IIB superstring action in AdS(5) x S(5) background,''
  Nucl.\ Phys.\ B {\bf 533}, 109 (1998)
  [hep-th/9805028].
  %%CITATION = HEP-TH 9805028;%%

\bibitem{RK}
  B.~Kulik and R.~Roiban,
  ``T-duality of the Green-Schwarz superstring,''
  JHEP {\bf 0209}, 007 (2002)
  [hep-th/0012010].
  %%CITATION = HEP-TH 0012010;%%
 M.~Cvetic, H.~Lu, C.~N.~Pope and K.~S.~Stelle,
  ``T-duality in the Green-Schwarz formalism, and the massless/massive
    IIA duality map,''
  Nucl.\ Phys.\ B {\bf 573}, 149 (2000)
  [hep-th/9907202].
  %%CITATION = HEP-TH 9907202;%%
  
  
\bi{BPR}
I.~Bena, J.~Polchinski and R.~Roiban,
  ``Hidden symmetries of the AdS(5) x S**5 superstring,''
  Phys.\ Rev.\ D {\bf 69}, 046002 (2004)
  [hep-th/0305116].
  %%CITATION = HEP-TH 0305116;%%


\bibitem{KMMZ}
V.~A.~Kazakov, A.~Marshakov, J.~A.~Minahan and K.~Zarembo,
``Classical / quantum integrability in AdS/CFT,'' JHEP {\bf 0405}
(2004) 024  hep-th/0402207.
%
\bibitem{BKSZ}
N.~Beisert, V.~A.~Kazakov, K.~Sakai and K.~Zarembo,
``The algebraic curve of classical superstrings on AdS(5) x S**5,''
hep-th/0502226.
%%CITATION = HEP-TH 0502226;%%

\bibitem{afs}
  G.~Arutyunov, S.~Frolov and M.~Staudacher,
  ``Bethe ansatz for quantum strings,''
  JHEP {\bf 0410}, 016 (2004)
  [hep-th/0406256].
  %%CITATION = HEP-TH 0406256;%%
%\bibitem{s}
M.~Staudacher,
``The factorized S-matrix of CFT/AdS,''
JHEP {\bf 0505}, 054 (2005)
[hep-th/0412188].
%%CITATION = HEP-TH 0412188;%%
%
%\bibitem{bs}
N.~Beisert and M.~Staudacher,
``Long-range PSU(2,2$|$4) Bethe ansaetze for gauge theory and strings,''
hep-th/0504190.
%%CITATION = HEP-TH 0504190;%%
%


\bi{STACK}
  N.~Beisert, V.~A.~Kazakov, K.~Sakai and K.~Zarembo,
  ``Complete spectrum of long operators in N = 4 SYM at one loop,''
  hep-th/0503200.
  %%CITATION = HEP-TH 0503200;%%

\bibitem{FrKr}
  L.~Freyhult and C.~Kristjansen,
  ``Rational three-spin string duals and non-anomalous finite size effects,''
  JHEP {\bf 0505}, 043 (2005)
  [hep-th/0502122].
  %%CITATION = HEP-TH 0502122;%%

\bibitem{BEFR}
  N.~Beisert and L.~Freyhult,
  ``Fluctuations and energy shifts in the Bethe ansatz,''
  hep-th/0506243.
  %%CITATION = HEP-TH 0506243;%%

%
%
%\bi{MUSH}
%  N.I.~Muskhelishvili,
%  ``Singular integral equations, boundary problems of function theory
%and their application to mathematical physics'',
% Groningen, P. Noordhoff, 1953.
%

\end{thebibliography}
\end{document}